\documentclass[amsthm]{elsart}

\usepackage{yjsco}
\usepackage{natbib}

\usepackage{graphicx}
\usepackage{algorithm}
\usepackage{algorithmic}
\usepackage[american]{babel}
\usepackage{tikz}
\usetikzlibrary{arrows,shapes,backgrounds}
\usepackage{multirow}
\usepackage{hhline}
\usepackage{xtab}
\usepackage{lscape}
\usepackage{rotating}
\usepackage{pgfplotstable}
\usepackage{color}
\usepackage{colortbl}
\usepackage{array} 
\usepackage{paralist} 
\usepackage{verbatim} 
\usepackage{subfigure} 
\usepackage[rflt]{floatflt}

\usepackage[T1]{fontenc}
\usepackage[utopia]{mathdesign}

\newcommand{\Bliss}{\addplot[mark=square*, mark options={draw=red!40!black, fill=red!90!yellow, scale=1, opacity=.74}, only marks]}
\newcommand{\Nauty}{\addplot[mark=star, mark options={scale=1.24, thick, draw=green!60!black, opacity=1}, only marks]}
\newcommand{\TracesOneD}{\addplot[mark=o, mark options={scale=1.2, thick, draw=blue!20!black, opacity=.74}, only marks]}
\newcommand{\TracesTwoD}{\addplot[mark=*, mark options={scale=1.2, very thin, draw=black,fill=blue, opacity=.74}, only marks]}
\newcommand{\TracesTwoDAut}{\addplot[mark=triangle*, mark options={scale=1.2, thin, draw=black,fill=yellow, opacity=.74}, only marks]}

\begin{document}
\begin{frontmatter}

\title{Search Space Contraction in Canonical Labeling of Graphs}


\author{Adolfo Piperno}
\address{Dipartimento di Informatica, La Sapienza Universit\`a di Roma, Via Salaria 113, I-00198 Roma}
\ead{piperno@di.uniroma1.it}
\ead[url]{http://www.dsi.uniroma1.it/$\sim$piperno/pers/Traces.html}

\begin{abstract}
The individualization-refinement paradigm for computing a canonical labeling 
and the automorphism group of a graph is investigated. A new
algorithmic design aimed at reducing the size of the associated search space is introduced, and 
a new tool, named \emph{Traces}, is presented, together with experimental results and comparisons
with existing software, such as McKay's \emph{nauty}.
It is shown that the approach presented here leads to a huge reduction in the 
search space, thereby making computation feasible for several classes of graphs which are 
hard for all the main canonical labeling tools in the literature.
\end{abstract}

\begin{keyword}
(Practical) graph isomorphism, canonical labeling, partition refinement, automorphism group computation. 
\end{keyword}

\end{frontmatter}
\newcommand{\delete}[1]{}

\delete{\section{Main Result}

We recall the works reported in \citep{Key1}.
......
We recall that \citet{Key1} proved the result.

\begin{defn}[Nice Notion]
We say a polynomial is \emph{nice} if ....  \qed
\end{defn}

We state the main result.

\begin{thm}[Someone 2005]
All nice polynomials are also pretty....  \qed
\end{thm}

\begin{pf}
The proof is easy... 
\end{pf}
}




\theoremstyle{plain}
\newtheorem{theorem}{Theorem}[section]
\newtheorem{lemma}[theorem]{Lemma}
\newtheorem{proposition}[theorem]{Proposition}
\newtheorem{property}[theorem]{Property}
\newtheorem{corollary}[theorem]{Corollary}
\newtheorem{problem}[theorem]{Problem}
\theoremstyle{definition}
\newtheorem{definition}[theorem]{Definition}
\newtheorem{notation}[theorem]{Notation}
\newtheorem{conjecture}[theorem]{Conjecture}
\newtheorem{example}[theorem]{Example}
\newtheorem{algo}[theorem]{Algorithm}
\theoremstyle{remark}
\newtheorem{remark}[theorem]{Remark}

\newcommand\per{{\hspace{.8pt}{\cdot}\hspace{.8pt}}}
\newcommand\T{\rule{0pt}{2.2ex}}
\newcommand\B{\rule[-1ex]{0pt}{0pt}}
\newcommand{\enne}{[{n}]}
\newcommand{\ennemeno}{\bar{n}}
\newcommand{\cano}{{\mathcal{C}}}
\newcommand{\tree}{{\mathcal{T}}}
\newcommand{\graphs}{\mathbb{G}}
\newcommand{\genne}{\mathbb{G}_{\enne}}
\newcommand{\partiz}{\Pi_{\enne}}
\newcommand{\partx}[1]{\Pi_{[{#1}]}}
\newcommand{\refind}[1]{\pi^{({#1})}}
\newcommand{\refpiind}[2]{{#1}^{({#2})}}
\newcommand{\insie}[1]{ \{#1\}}
\newcommand{\pos}{\emph{pos\,}}
\newcommand{\ind}{\emph{ind\,}}
\newcommand{\posc}{{\mathit pos}}
\newcommand{\raff}[1]{\varrho(#1)}
\newcommand{\pair}[2]{\langle#1,#2\rangle}
\newcommand{\AttSet}{\pmb{CandSet}}
\newcommand{\IndiSet}{\pmb{IndiSet}}
\newcommand{\DiscSet}{\pmb{DiscSet}}
\newcommand{\Var}[1]{\mathit{#1}}
\newcommand{\Filter}{{\tt Filter}}
\newcommand{\AutomFilter}{{\tt AutomFilter}}
\newcommand{\menospazio}{\vspace{-6pt}}
\newcommand{\att}[1]{{\mathbb #1}}
\newcommand{\indi}[1]{\pi{\downarrow}#1}
\newcommand{\indiA}[2]{#1{\downarrow}#2}
\newcommand{\indibar}[1]{\overline{\pi}{\downarrow}#1}
\newcommand{\indivar}[2]{#1_{[#2]}}
\newcommand{\BBCAN}{{\sl BBCAN}}
\newcommand{\allarga}[1]{\makebox[8pt][c]{{\scriptsize\texttt{#1}}}}
\newcommand{\indici}[1]{\multicolumn{1}{@{}c@{}}{\tiny #1}}
\newcommand{\text}[1]{\textrm{#1}}

\definecolor{Col0}{rgb}{0.70,0.02,0.00}
\definecolor{Col1}{rgb}{0.02,0.70,0.02}
\definecolor{Col2}{rgb}{1.00,0.75,0.00}
\definecolor{Col3}{rgb}{0.15,0.75,0.75}
\definecolor{Col4}{rgb}{0.20,0.20,0.70}
\definecolor{Col5}{rgb}{0.70,0.05,0.70}
\definecolor{Col6}{rgb}{0.70,0.40,0.02}
\definecolor{Col7}{rgb}{1.00,0.45,0.00}
\definecolor{Col8}{rgb}{0.90,0.70,0.80}
\definecolor{Col9}{rgb}{0.75,0.75,0.75}
\definecolor{Col10}{rgb}{0.70,0.90,0.70}
\definecolor{Col12}{rgb}{0.65,0.65,0.90}
\definecolor{Col16}{rgb}{0.90,0.80,0.70}
\definecolor{Col18}{rgb}{0.50,0.50,0.50}

\newcommand{\cancella}[1]{}
\newcommand{\CANCELLA}[1]{}

\pgfdeclareradialshading{sphereCol0}{\pgfpoint{2pt}{2pt}}%
{rgb(0pt)=(1,0.6,0);
rgb(2.4pt)=(0.7,0,0);
rgb(4pt)=(0.5,0,0)
}
\pgfdeclareradialshading{sphereCol1}{\pgfpoint{2pt}{2pt}}%
{rgb(0pt)=(.6,1,.6);
rgb(2.4pt)=(0,0.7,0);
rgb(4pt)=(0,0.5,0)
}
\pgfdeclareradialshading{sphereCol2}{\pgfpoint{2pt}{2pt}}%
{rgb(0pt)=(1,1,0);
rgb(2.4pt)=(1,0.7,0);
rgb(4pt)=(1,0.5,0)
}
\pgfdeclareradialshading{sphereCol3}{\pgfpoint{2pt}{2pt}}%
{rgb(0pt)=(0.6,1,1);
rgb(2.4pt)=(0,0.7,0.7);
rgb(4pt)=(0,0.5,0.5)
}
\pgfdeclareradialshading{sphereCol4}{\pgfpoint{2pt}{2pt}}%
{rgb(0pt)=(0.6,0.6,1);
rgb(2.4pt)=(0.2,0.2,0.7);
rgb(4pt)=(0.2,0.2,0.5)
}
\pgfdeclareradialshading{sphereCol5}{\pgfpoint{2pt}{2pt}}%
{rgb(0pt)=(1,0.6,1);
rgb(2.4pt)=(0.7,0,0.7);
rgb(4pt)=(0.5,0,0.5)
}
\pgfdeclareradialshading{sphereCol6}{\pgfpoint{2pt}{2pt}}%
{rgb(0pt)=(1,0.7,0.2);
rgb(2.4pt)=(0.7,0.4,0);
rgb(4pt)=(0.5,0.2,0)
}
\pgfdeclareradialshading{sphereCol7}{\pgfpoint{2pt}{2pt}}%
{rgb(0pt)=(1,0.8,0);
rgb(2.4pt)=(1,0.5,0);
rgb(4pt)=(1,0.3,0)
}
\pgfdeclareradialshading{sphereCol8}{\pgfpoint{2pt}{2pt}}%
{rgb(0pt)=(1,0.9,0.9);
rgb(2.4pt)=(0.9,0.7,0.8);
rgb(4pt)=(0.7,0.5,0.6)
}
\pgfdeclareradialshading{sphereCol9}{\pgfpoint{2pt}{2pt}}%
{rgb(0pt)=(0.9,0.9,0.9);
rgb(2.4pt)=(0.7,0.7,0.7);
rgb(4pt)=(0.5,0.5,0.5)
}

\pgfdeclareradialshading{sphereCol10}{\pgfpoint{2pt}{2pt}}%
{rgb(0pt)=(1,1,1);
rgb(2.4pt)=(0.7,0.9,0.7);
rgb(4pt)=(0.5,0.9,0.5)
}
\pgfdeclareradialshading{sphereCol11}{\pgfpoint{2pt}{2pt}}%
{rgb(0pt)=(1,1,1);
rgb(2.4pt)=(0.7,0.7,0.9);
rgb(4pt)=(0.5,0.5,0.9)
}
\pgfdeclareradialshading{sphereCol12}{\pgfpoint{2pt}{2pt}}%
{rgb(0pt)=(1,1,1);
rgb(2.4pt)=(0.9,0.8,0.7);
rgb(4pt)=(0.9,0.8,0.5)
}
\pgfdeclareradialshading{sphereCol13}{\pgfpoint{2pt}{2pt}}%
{rgb(0pt)=(0.8,0.8,0.8);
rgb(2.4pt)=(0.5,0.5,0.5);
rgb(4pt)=(0.2,0.2,0.2)
}

\newcommand{\palla}[3]{\draw #2 circle (4pt);
\draw #2 node {\pgfuseshading{#1}} node[above=6pt,left=1pt] {\scriptsize $#3$};}
\newcommand{\pallaright}[3]{\draw #2 circle (4pt);
\draw #2 node {\pgfuseshading{#1}} node[above=6pt,right=1pt] {\scriptsize $#3$};}

\section{Introduction}
A \emph{canonical labeling} (or  \emph{canonical form}) 
of a graph $G$ is a graph $G'$ --- isomorphic to $G$ --- representing the whole isomorphism class of $G$.
In terms of computational complexity the theoretical status of canonical labeling (CL) is still unsettled, 
since an efficient algorithm for  CL would imply an efficient algorithm
for the graph isomorphism problem (GI). In practice, however, CL algorithms 
are widely used, as they enable (possibly large) sequences of graphs coming from both
combinatorial problems and industrial applications to be checked
for isomorphism by simply comparing their canonical forms.

The literature on methods for approaching GI and CL displays a peculiar ``separation''
between theoretical and practical studies.
On the theoretical side, besides papers substantiating the thesis that GI is 
not NP{-}complete \cite{Mathon79, Schoning88} (a survey is in \cite{ArvindT05}),
there are a large number of noteworthy pieces of mathematics showing the existence of polynomial
solutions of GI for significant classes of graphs. While moderately exponential solutions have been provided for
the general problem of graph isomorphism \cite{Babai81, BabaiL83}, 
polynomial algorithms exist for planar graphs \cite{HopcroftT73, HopcroftW74}, graphs of
bounded genus \cite{FilottiM80}, graphs with colored vertices and bounded color-classes \cite{Babai79}, graphs with bounded
multiplicity of eigenvalues \cite{BabaiGM82}, graphs of bounded valence \cite{Luks82}, and more (see \cite{BabaiHand}). 

On the practical side, there are some noteworthy pieces of software which originate from the outstanding tool
\emph{nauty} \cite{nauty22}. \emph{nauty} was introduced in the 1980s by 
McKay \cite{McKay81} and has become a standard in the area of 
canonical labeling and determination of the automorphism group of a graph.
Moreover, it has been incorporated into more general mathematical software tools such as \texttt{GAP} \cite{gap}
and \texttt{MAGMA} \cite{magma}.

It is important to observe that, with the exception of planar graphs, none of the polynomial algorithms
mentioned above has been implemented in software, as noted by 
Junttila and Kaski in \cite{JunttilaK07}. A reasonable justification for this absence would seem to be that,
given a class $\mathbb{C}$ of graphs for which an efficient algorithm for isomorphism testing exists, 
\emph{nauty} is usually able to process almost all the graphs in $\mathbb{C}$ in a 
considerably smaller number of steps than that established by the theoretical bound (with respect to the number of vertices).  
However, there exist graphs in $\mathbb{C}$ for which \emph{nauty} exhibits an exponential behavior, 
as shown by the series of graphs constructed by Miyazaki in \cite{miyazaki}: 
all these graphs are $3$-regular and have color-class size equal to $4$, 
hence they intersect two classes of graphs for which polynomial solutions for GI exist. 
A further distinctive feature of Miyazaki's graphs is that the size of their automorphism group is quite large. 
This contrasts with the fact that graphs which are hard for \emph{nauty} usually
have a high degree of regularity but a small automorphism group.

In recent years the tools \emph{saucy} \cite{DargaLSM04,saucy} and 
\emph{bliss} \cite{JunttilaK07,bliss} have been introduced, aimed at
handling large sparse graphs coming either from the satisfiability problem (SAT), or from industrial applications. 
Like \emph{nauty} these are general purpose devices implementing backtrack algorithms based on the so called 
\emph{individualization-refinement} technique, however they differ from \emph{nauty} in respect of the 
data structures and heuristics used. 
We briefly recall here that the key of the individualization-refinement technique is the notion
of \emph{equitable partition}, a coloring of vertices of a graph such that any two vertices with the same color
have the same number of neighbors in each color class.
A vertex is \emph{individualized} by assigning to it a fresh color, the coloring so obtained is \emph{refined} 
when a new equitable partition (finer than the initial one) is produced.
A backtrack search on the space of all possible individualizations, along with some initial assumptions
that allow the associated tree to be pruned, produces the canonical form for the input graph.
The complexity of a tool based on such a technique essentially derives from 
the size of the associated search space, and from the complexity of the refinement
function (which is invoked once for every node of the search tree). On the other hand, the granularity of the refinement 
procedure (see \emph{nauty}'s \emph{vertex-invariants} \cite{nauty22}) may affect the size of the search tree. 


In the present paper we start our investigation from an analysis of the main critical features of the individualization-refinement paradigm,
namely (i) the visiting strategy of the search space, (ii) the manipulation of discovered automorphisms of the input graph, 
(iii) the refinement function, and (iv) the selection of the next individualization step.
If we disregard implementation details, all four of these issues are treated in essentially the same way by each of the
existing tools for canonical labeling of graphs, revealing that they are based on the same algorithmic conception 
--- that introduced by \emph{nauty} and described above. We will, instead, propose a new solution for each of the four issues, presenting
a canonical labeling tool, named \emph{Traces}, whose algorithmic structure differs from that  of all previous tools.
In fact, it is clear from the comparison between \emph{nauty} and \emph{bliss} that the adoption of suitable heuristics and
data structures can produce a (sometimes significant) improvement in performance. But classes of graphs that are very hard
for \emph{nauty} are still very hard for \emph{bliss}: this seems to be due to the fact that these tools
have the same conceptual bases. Therefore, it is our aim to experiment a tool with a different basic design. 

\cancella{ of the individualization-refinement paradigm 
for canonical labeling of graphs: the size and depth of the search tree, the refinement procedure, the selector of
the next individualization step. We suggest a different structure for the refinement procedure, which will be
called \emph{multi-refinement}, whose corresponding invariant is not based on local properties of vertices of the graph
(such as adjacencies, incident cliques and others): 
two vertices will be considered equivalent (and therefore assigned to the same color class) when their individualizations induce
(after refinement) equivalent partitions. 
This enables to reduce the depth of the search tree, obtaining finer partitions, and authorizes 
new criteria for selecting the color class of the next 
individualized vertices, in our case the one producing the partition with maximal size.

Our multi-refinement is computationally much heavier than usual refinement. In the worst case, we could
be losing a factor of $n$ (the number of vertices of the considered graph) in time, when the refinements of all vertex
individualizations are needed to produce the multi-refinement. However, from one hand some techniques will be introduced 
in order to manage the required refinements without computing them completely, including the use of 
information coming from the automorphism group of the graph and a new representation of the whole refinement process.
From the other hand, multi-refinement naturally suggests its parallel execution, though this is outside the scope of the present paper.

On the positive side, multi-refinement will cause a considerable reduction of the size of the search tree:
in particular, such decrease will be shown to be dramatic when very large search trees are needed by \emph{nauty} and other tools;
as an example, an exponential reduction of the size of the search tree will come out for Miyazaki's sequence of graphs.
In all cases, multi-refinement, combined with the associated selector of the next individualization step, 
causes the reduction of the depth of the search tree. We observe that the contraction of one level of the
search tree often pays back, in terms of time, the loss caused by multi-refinement.}

\emph{Traces}\footnote{\emph{Traces} is available at
{\tt http://www.dsi.uniroma1.it/$\sim$piperno/pers/Traces.html}} computes a canonical form for a 
colored graph and/or a set of generators for its automorphism group.
Its main innovations of  can be summarized as follows: 
\emph{Traces} does not use backtrack to traverse the search space: the case for adopting a kind of breadth-first strategy will
be argued in this paper. 
Automorphisms which are detected are manipulated in \emph{Traces} by means of the Schreier-Sims algorithm \cite{sims}
(see also \cite{Seress02}); information about the group structure can be also used by the refinement procedure
to eliminate redundant computations. 
In addition to the usual one, a refinement function producing finer partitions can be used for difficult graphs; such function
also provides information for choosing the next individualization step. 
Partitions are compared (and possibly discarded) without computing them completely, using a linear representation which we call a
\emph{trace}. 

We produce performance tables using the huge catalogue of benchmark graphs for
canonical labeling and automorphism group computation compiled by Junttila and Kaski \cite{JunttilaK07}. 
\cancella{In order to give an anticipation of our results and a final account on our motivations, we report some experiments, 
taken from Tables \ref{t:res} and \ref{t:res2} and executed on an Apple Mac Pro with 2 Quad Core processors at 2.8 GHz:
\begin{compactitem}
\item[-] \textsf{pp-16-9} is a $(546\textrm{ vertices},4641\textrm{ edges})$ graph associated to one of the 
known projective planes of order $16$ (see \cite{RoylePlanes}): \emph{bliss}, which is 
considerably faster than \emph{nauty} on this graph, takes approximately one hour to canonize it, traversing more than $300$ million nodes of the search tree, which has
depth $6$; \emph{Traces} reduces the depth of the search tree down to $3$ and traverses $187$ nodes, only, in $0.39$ seconds; the time spent in group computation
is in this case negligible; the size of the automorphism group of the graph is $921,600$;
\item[-] \textsf{had-100} is a $(400,20200)$ graph associated to a $100\times100$ Hadamard matrix (see \cite{HadaSloane}): \emph{bliss}
takes $4.23$ seconds to canonize \textsf{had-100}, traversing more than $50,000$ nodes of the search tree, which has
depth $5$; \emph{Traces} reduces the depth of the search tree down to $2$ and traverses $725$ nodes in $3,20$ seconds; the time spent in group computation
is in this case negligible, too; the size of the automorphism group of the graph is $400$;
\item[-] while canonizing a complete graph, \emph{Traces} spends almost all the time in group computation;
\item[-] the sequence of graphs \textsf{mz-aug2-2x} derives from Miyazaki's construction; \emph{bliss}
traverses more than $1$ million nodes to canonize \textsf{mz-aug2-18} (a $(432,684)$ graph), 
more than $4$ million nodes to canonize \textsf{mz-aug2-20} (a $(480,760)$ graph),
more than $16$ million nodes to canonize \textsf{mz-aug2-22} (a $(528,836)$ graph), and so on. 
On its hand, \emph{Traces} traverses $145$ nodes to canonize \textsf{mz-aug2-18}, 
$161$ nodes to canonize \textsf{mz-aug2-20}, $177$ nodes to canonize \textsf{mz-aug2-22}, a linear
sequence.
\end{compactitem}
}
It turns out that, compared to other tools, \emph{Traces} is able to significantly reduce the size of the search space 
in the case of hard graphs, often by several orders of magnitude. When the ratio between sizes of the search spaces
is substantial, then \emph{Traces} runs much faster than any other tool; otherwise, their timings are mostly comparable. 
Even better results are obtained in the case of automorphism group computation instead of canonical labeling.
\emph{Traces} may be slightly slower than some of the other tools when the associated search spaces are very small, usually when
the input graph has a low degree of regularity or a high degree of symmetry. This is mainly due to the absence (in the current
version) of a refinement procedure specific for sparse graphs.

All known classes of graphs which are intractable by the state-of-the-art tools are efficiently treated by \emph{Traces},
though we need to point out that some of these classes were carefully tuned to cause those tools to behave exponentially.
Particular attention will be given below to incidence graphs of projective planes, which are considered to be the hardest instances
for canonical labeling (see discussions at \emph{nauty} mailing list
\footnote{\texttt{http://dcsmail.anu.edu.au/cgi-bin/mailman/listinfo/nauty-list}}). 
We will show that \emph{Traces}, which is a general purpose device, treats these graphs within the best
theoretically established bound, without the help of any additional procedure tailored to them. 

\subsection{Structure of the paper}
In Section \ref{s:grapart} some definitions and properties of graphs and partitions are introduced and
the individualization-refinement technique is described. 
Section \ref{s:mot} presents an analysis of some fundamental issues in canonical labeling and argues the need for
a new algorithmic design.
Section \ref{s:canlabalg} is dedicated to the new canonical labeling algorithm ant to its correctness proof. 
Experimental results are presented and commented upon in section \ref{s:exper}.

\section{Graphs and partitions}\label{s:grapart}


A (simple) {\em graph} is a pair $(V,E)$, where $V$ is a finite set of
\emph{vertices} and $E$ is a set of unordered pairs of vertices called {\em edges}.
If $(u,v)\in E$, we say that $u$ and $v$ are {\em adjacent} or {\em neighbors}.
A (vertex) {\em colored graph} is a pair $G=(H,\chi)$, where $H$ is a graph and
$\chi$ is a function assigning colors to vertices of $H$. 

In this paper, $\enne$ will denote the set $\{1,\dots,n\}$, while $\genne$
will denote the set of colored graphs with vertex set $\enne$. 
The set of all colored graphs will be denoted by $\graphs$.
Note that any graph is a colored graph in which all 
vertices have the same fixed color, and that it is not restrictive to assume
that $\chi : \enne\rightarrow\enne$.

An {\em isomorphism} of graphs $G_1,G_2\in\genne$ is a permutation $p$ of $\enne$ 
such that any two vertices $u$ and $v$ of $G_1$ are adjacent in $G_1$ if and only if $p(u)$ and $p(v)$ are adjacent in $G_2$. 
When considering graphs with colored vertices, isomorphisms must preserve colors, too.
We will write $G_1\simeq G_2$ when $G_1$ and $G_2$ are isomorphic. 
An \emph{automorphism} is an isomorphism between a graph and itself.
The \emph{automorphism group} $\textrm{Aut}(G)$ of a graph $G$ is the set of all automorphisms of $G$ with
permutation composition as the group operation. If $\Gamma\subseteq\textrm{Aut}(G)$ is a group of automorphisms
of $G$, then $\Gamma$ induces an equivalence relation on the vertices of $G$: two vertices
$v,w$ are equivalent if and only if there exists an automorphism in $\Gamma$ which maps
$v$ to $w$; the resulting equivalence classes are called the \emph{orbits} of the graph by $\Gamma$. 

%
A function $f$ from $\graphs$ to a set $\mathbb{D}$ is an {\em (isomorphism) invariant} 
on $\graphs$ iff $\forall G_1,G_2\in\graphs:\,G_1\simeq G_2 \Rightarrow f(G_1) = f(G_2).$
The image of a graph $G$ under a function $\cano:\graphs\rightarrow\graphs$ is a \emph{canonical labeling}
(or \emph{canonical form}) of $G$ iff
(i) $\forall G\in\graphs:\,\cano(G)\simeq G$,
(ii) $\forall G_1,G_2\in\graphs:\,G_1\simeq G_2~\Leftrightarrow~\cano(G_1)=\cano(G_2)$.

An \emph{ordered partition} of $\enne$ is a sequence $\pi=(W_1,\dots,W_r)$ of disjoint
non-empty subsets of $\enne$, called \emph{cells}, whose union is $\enne$. 
The set of ordered partitions of $\enne$ will be denoted by $\partiz$, while $\Pi$ will denote
the set of all ordered partitions. The \emph{size} of a partition is the number of its cells

A cell of a partition $\pi\in\partiz$ is \emph{trivial} when it contains only one
element. The partition $\pi$ is \emph{discrete} if all its cells are trivial; $\pi$ is the 
\emph{unit partition} when it has only one cell, i.e.\ $\pi=(\enne)$. 
For any $\pi\in\partiz$ and $v,w\in\enne$, we will write $v\sim_\pi w$ when $v$ and $w$
belong to the same cell of~$\pi$.
An \emph{orbit partition} of a graph $G$ with respect to a subgroup $\Gamma$ of $\textrm{Aut}(G)$
is any partition of vertices of $G$, whose cells are the orbits of $G$ under $\Gamma$. 


The \emph{index} $\ind(v,\pi)$ of a vertex $v\in\enne$ in an ordered partition $\pi\in\partiz$ is 
the index of the cell of $\pi$ in which $v$ appears, namely 
$\ind(v,(W_1,\dots,W_r)) = k$ when $v\in W_k$.
The \emph{position} of a vertex $v\in\enne$ in an ordered partition $\pi\in\partiz$
is defined by means of the function $\pos:\enne\times\partiz\rightarrow\enne$ such that
$ \ind(v,(W_1,\dots,W_r)) = k \Rightarrow \pos(v,(W_1,\dots,W_r)) = 1+\sum_{i=1}^{k-1}\mid W_i\mid$.
The position of a cell $W$ in an ordered partition $\pi$ is defined as the position of 
an element of $W$ in $\pi$ (indeed, all the elements of $W$ share the same position in $\pi$); 
with some overloading: $\pos(W,\pi) = \pos(v,\pi), \mbox{ for any $v\in W$}$.

For example, if $\pi=(\{2,3\},\{5\},\{1,4\})\in\partx{5}$, then 
$\pos(2,\pi)=\pos(3,\pi)= 1 =\pos(\insie{2,3},\pi)$, while 
$\pos(1,\pi)=\pos(4,\pi)= 4 = \pos(\insie{1,4},\pi)$.
In particular, $\pos(v,(\enne))=1$, for any $v\in\enne$.

If $\pi_1$ and $\pi_2$ are partitions, then $\pi_1$ is \emph{finer} than $\pi_2$, 
and $\pi_2$ is  \emph{coarser} than $\pi_1$, if every cell of $\pi_1$ is a subset 
of some cell of $\pi_2$. (Note that partitions are both
finer and coarser than themselves, and that $\pi_1$ is finer than $\pi_2$ iff
$\forall v\in\enne:\, \pos(v,\pi_1)\geqslant\pos(v,\pi_2)$.)

Let $\pi\in\partiz$ and let $(\chi_1,\dots,\chi_n)$ be a sequence of $n$ colors. 
The graph $G=(H,\pi)$ is a colored graph if we interpret the partition $\pi$ as a function 
assigning the $\pos(v,\pi)$-th color in $(\chi_1,\dots,\chi_n)$ to vertex $v$ of $G$. 
We observe that the converse is also true: an ordered sequence of colors induces an
ordered partition of vertices of a colored graph. 
Therefore in the rest of the paper we will denote a colored graph as a pair 
$G=(H,\pi)$, where $\pi$ is an ordered partition of vertices of $H$, implicitly 
assuming the existence of an ordered sequence $(\chi_1,\dots,\chi_n)$ of 
colors and a function $\chi$ such that, for any vertex $v$, $\chi(v)=\chi_{\pos(v,\pi)}$.


\subsection{The individualization-refinement technique for canonical labeling}\label{s:indref}

In this section the behavior of the individualization-refinement technique for canonical labeling
is revisited. The method consists of a depth-first search of a space defined by classification 
and fixing of vertices. It is based on the fact that any permutation $\gamma$ of vertices of a graph
$G$ can be viewed as a discrete partition whose cells appear in the order established by $\gamma$.
Vertices of $G$ are partitioned by a \emph{refinement function}, which separates them in a way that
no automorphism of $G$ exists between different cells. The \emph{individualization}
of a vertex corresponds to select a subgroup of permutations fixing that vertex. 

Let $\pi=(W_1,\dots,W_r)\in\partiz$ be an ordered partition of $\enne$.
For any $v\in\enne$, if $v$ belongs to a non-trivial cell $W_i$, then we denote by $\indi{v}$ the 
ordered partition obtained from $\pi$ 
by splitting the cell $W_i$ into the cells $\insie{v}$ and $W_i-\insie{v}$, namely
$\indi{v} = (W_1,\dots,W_{i-1},\insie{v},W_i-\insie{v},W_{i+1},\dots,W_r), \mbox{ if $v\in W_i$}$.

If $G=(H,\pi)\in\genne$, then we say that $G'=(H,\indi{v})$ is obtained from $G$ by 
\emph{individualizing} vertex $v$.
Note that 
$\pos(w,\indi{v}) =
\pos(w,\pi)+1 \mbox{ if $w\in W_i-\insie{v}$.}$

Let $G=((V,E),\pi)\in\graphs$ be a colored graph. 
We say that $G$ is \emph{equitable} when, for any pair of 
vertices $v,w\in V$, if $v\sim_\pi w$, then, for any cell $W$ of $\pi$, $v$ and $w$ 
have the same number of neighbors in $W$.
In this case, we say that $\pi$ is a \emph{stable} partition for $(V,E)$.

Let $G=((V,E),\pi)\in\graphs$ be an equitable graph. The \emph{quotient graph} $Q(G)=(V',E')$ of $G$ 
is a graph, with possible multiple edges and loops, having vertex set $V' = \{\pos(v,\pi)\mid v\in V\}$
and edge multiset $E'=\{\!\!\{(\pos(v,\pi),\pos(w,\pi))\mid(v,w)\in E\}\!\!\}$.

\begin{figure}[h]
\centering
\includegraphics[scale=.22]{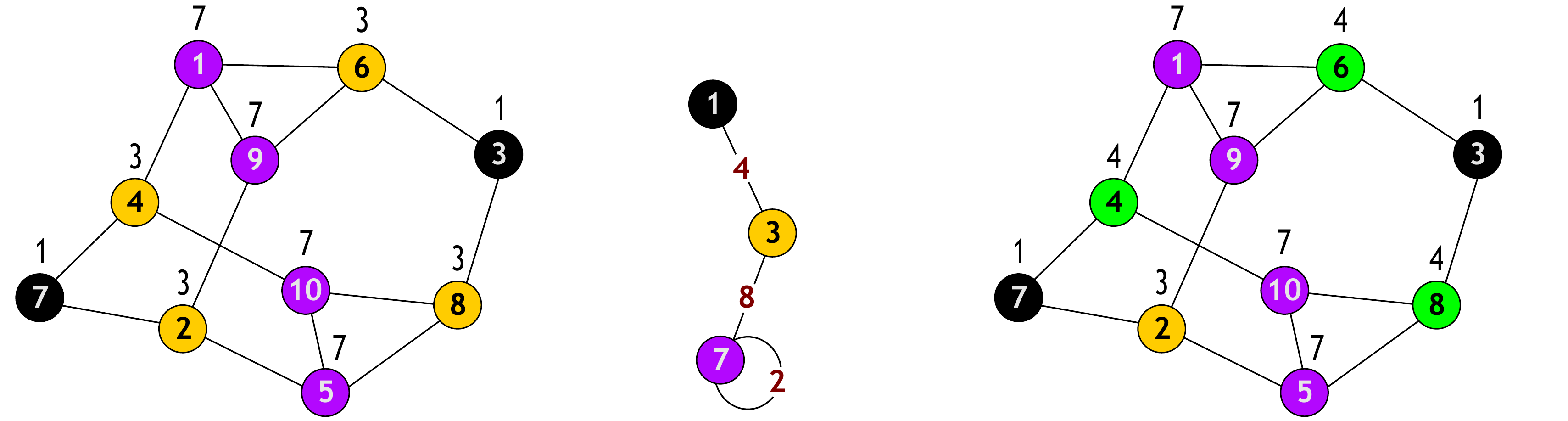}
\caption{An equitable graph (left) and its quotient graph (center); individualization of vertex $2$ (right). Numbers above vertices denote colors, which correspond to vertices of the quotient graph.}
\label{f:G10}
\end{figure}

Figure \ref{f:G10} shows an equitable graph $G$ (left) and its quotient graph (center). 
Figure \ref{f:G10} (right) displays the effect of individualizing vertex $2$. The individualized
vertex keeps its color ($3$), while other vertices in the original class take the next color ($4$). 
For every color appearing in $G$, there exists a corresponding vertex in $Q(G)$; an edge in $Q(G)$
has multiplicity $k$ if $G$ has $k$ edges joining vertices with the corresponding colors.

Note that when the coloring partition of a graph $G$ is discrete, then $Q(G)$ is isomorphic to $G$ itself.
Hence, if the quotients of two graphs with discrete color partitions are
the same, then the two graphs are isomorphic.

Let $G=(H,\pi)\in\graphs$. A \emph{refinement} of $G$ is the image of $G$ under a function 
$R:\graphs\rightarrow\graphs$ such that: 
(i) $R(H,\pi)=(H,\pi')$ is an equitable graph with $\pi'$ finer than $\pi$; 
(ii) $R$ preserves isomorphisms, i.e.\ 
$(H_1,\pi_1)\simeq (H_2,\pi_2) \Rightarrow R(H_1,\pi_1)\simeq R(H_2,\pi_2)$.
To simplify the notation, when considering a function $F$ on $\mathbb{G}$ we will write $F(H,\pi)$ instead of $F((H,\pi))$.

The refinement procedures implemented in canonical labeling tools (see also \cite{kocay})
can be briefly described as follows. Using a specific scheduling, which is isomorphism invariant, 
cells of $\pi$ are visited. For each visited cell $W$, the multiplicities of adjacencies of elements of $W$
are counted. The vertices in each cell $Z$ are divided into
subcells according to how many neighbors they have in $W$. This process is repeated until
the graph is equitable. 

\begin{figure}[htbp]
\centering
\includegraphics[scale=.21]{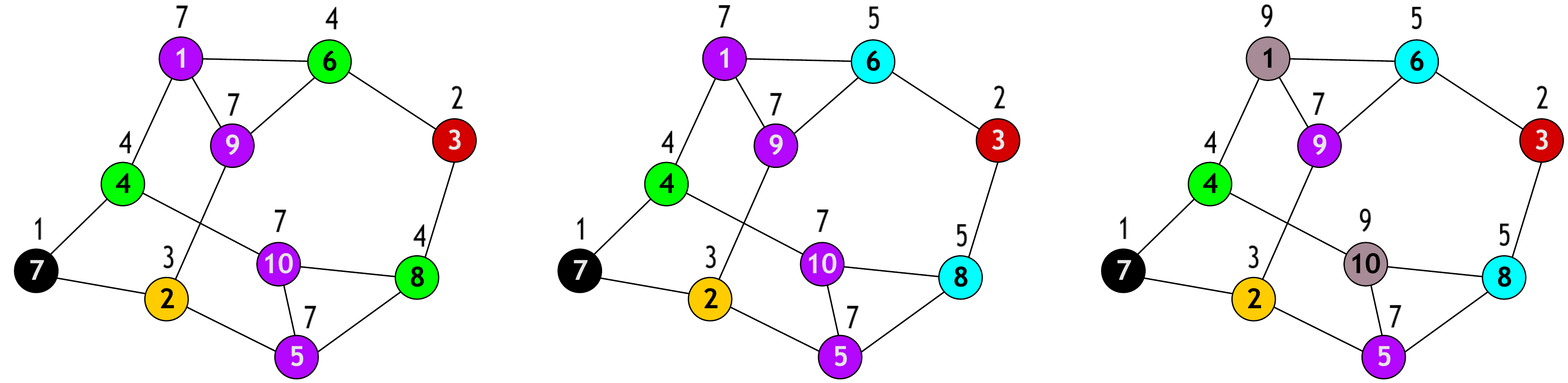}
\caption{Refinement}
\label{f:G10-1}
\end{figure}

In Figure \ref{f:G10-1}, starting from the non-equitable
colored graph in Figure \ref{f:G10} (right), the cell $\insie{3,7}$ is split 
(Figure \ref{f:G10-1} (left)), since vertex $7$, but not $3$, has one neighbor in the cell $\insie{2}$. 
Similarly, the cell $\insie{4,6,8}$ is split (Figure \ref{f:G10-1} (center)), since
vertex $4$, but not $6$ and $8$, has one neighbor in the cell $\insie{7}$. 
Finally, the cell $\insie{1,5,9,10}$ is split (Figure \ref{f:G10-1} (right))
since vertices $5$ and $9$, but not $1$ and $10$, have one neighbor in the cell $\insie{2}$.
The graph in Figure \ref{f:G10-1} (right) is equitable.

The canonical labeling algorithm implemented by \emph{nauty}, \emph{saucy} and \emph{bliss} 
is a depth-first search of a space defined by partition refinement and vertex individualization. 
Vertices are individualized according to a \emph{target cell selector}. 
This is a function $T:\graphs\rightarrow\enne$ such that: 
(i) $T(H,\pi)=k$ is the position of a non-trivial cell of $\pi$; 
(ii) $T$ is an isomorphism invariant, i.e.\ 
$(H_1,\pi_1)\simeq (H_2,\pi_2) \Rightarrow T(H_1,\pi_1)= T(H_2,\pi_2)$.
The image of a graph $G$ under $T$ is the \emph{target cell} of~$G$.

The target cell selector used in \emph{nauty}, \emph{saucy} and \emph{bliss} is defined in
terms of adjacencies between cells, which are classified according to the presence of edges and non-edges
with respect to other cells. The leftmost cell having the highest value in such classification 
is chosen as the target cell.

Given an equitable graph $G=(H,\pi)$, a refinement function $R$, and a non-empty sequence
 $v_1,\dots,v_k$ of vertices of $H$, 
we will write $(H,\refind{v_1,\dots,v_k})$ to denote the graph obtained by individualization 
and refinement of $v_1,\dots,v_k$, consecutively; more precisely
$$(H,\refind{v_1}) = R(H,\indi{v_1}) \mbox{ and }
(H,\refind{v_1,\dots,v_k}) = R(H,\indiA{\refind{v_1,\dots,v_{k-1}}}{v_k}).$$

\subsection{Backtrack construction of the search tree}

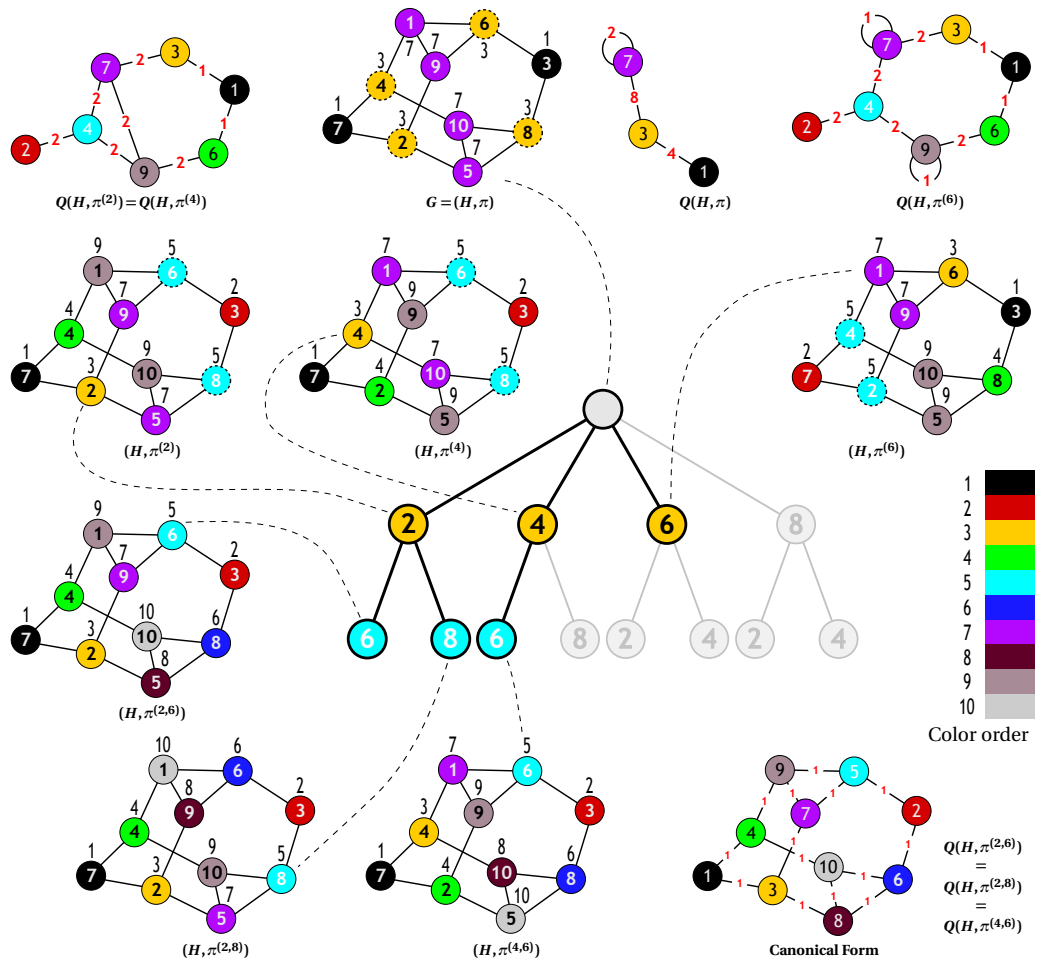
\begin{figure}[ht]
\begin{picture}(384,364)
\centering
\includegraphics[scale=.22]{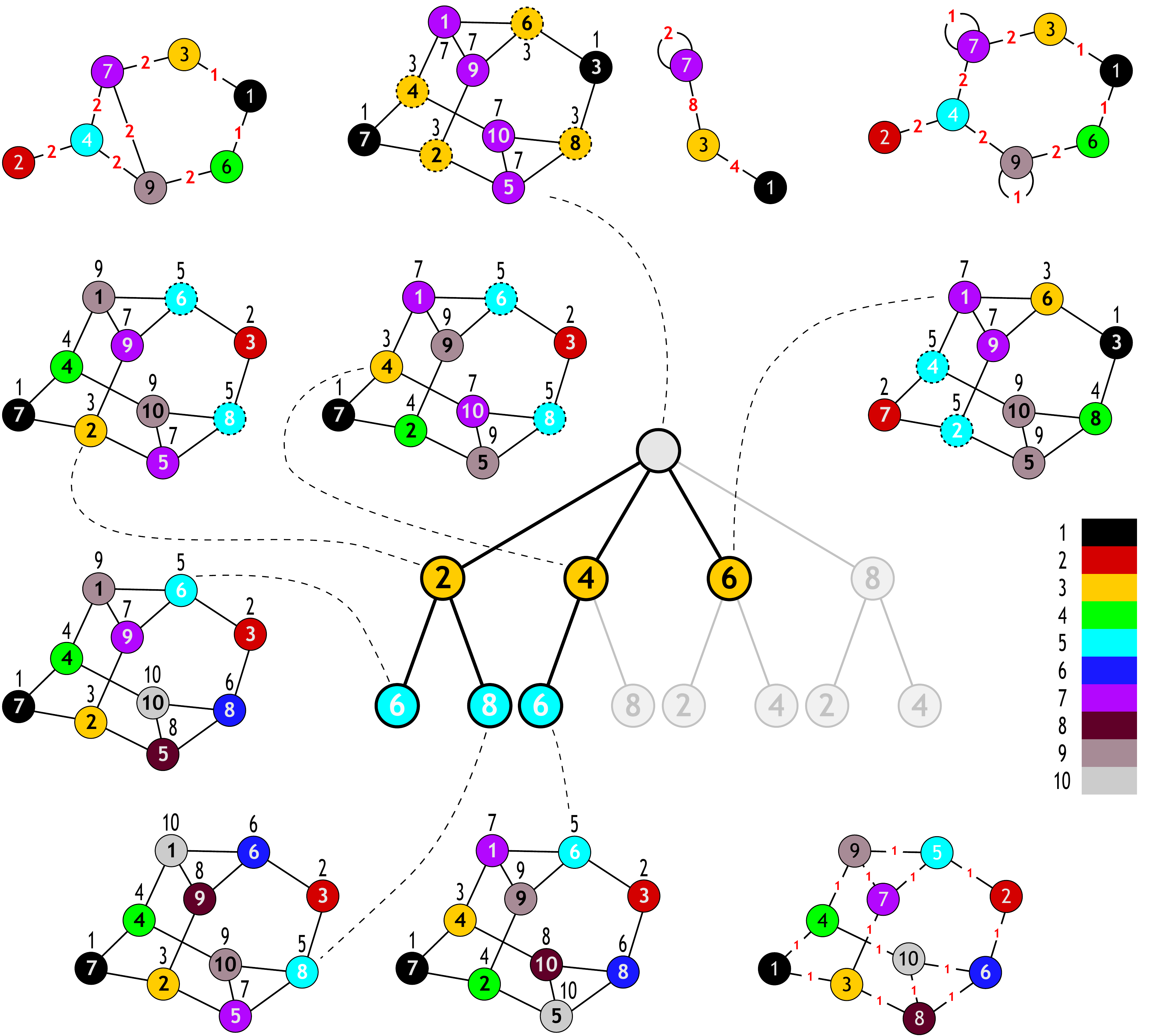}
\put(-238,276){\boldmath{\tiny$G=(H,\pi)$}}
\put(-142,276){\boldmath{\tiny$Q(H,\pi)$}}
\put(-378,276){\boldmath{\tiny$Q(H,\refind{2})=Q(H,\refind{4})$}}
\put(-60,276){\boldmath{\tiny$Q(H,\refind{6})$}}
\put(-352,182){\boldmath{\tiny$(H,\refind{2})$}}
\put(-242,182){\boldmath{\tiny$(H,\refind{4})$}}
\put(-78,182){\boldmath{\tiny$(H,\refind{6})$}}
\put(-354,82.6){\boldmath{\tiny$(H,\refind{2,6})$}}
\put(-330,-7.2){\boldmath{\tiny$(H,\refind{2,8})$}}
\put(-220,-7.2){\boldmath{\tiny$(H,\refind{4,6})$}}
\put(-42,32){\boldmath{\tiny$Q(H,\refind{2,6})$}}
\put(-30,25){\boldmath{\tiny$=$}}
\put(-42,17){\boldmath{\tiny$Q(H,\refind{2,8})$}}
\put(-30,10){\boldmath{\tiny$=$}}
\put(-42,2){\boldmath{\tiny$Q(H,\refind{4,6})$}}
\put(-108,-7.2){\tiny\bf Canonical Form}
\put(-48,74){\scriptsize Color order}
\end{picture}
\caption{Backtrack construction and pruning of the search tree}\label{f:searchtree}
\end{figure}

Assuming a fixed refinement function $R$ and a target cell selector $T$,
the {\em search tree} $\tree(H,\pi)$ of a graph $G=(H,\pi)$ is a tree in which each node
represents an equitable graph $G'=(H,\pi')$ where $\pi'$ is finer than $\pi$; in particular:%
\begin{compactitem}
\item[-] the root of $\tree(H,\pi)$ represents the refinement of $G$;
\item[-] if a node of $\tree(H,\pi)$ represents $(H,\pi')$ and $\pi'$ is discrete, 
then that node is a leaf;
\item[-] otherwise, let $\pi_\nu = (W_1,\dots,W_i,\dots,W_k)$ be the partition of the
graph $(H,\pi_\nu)$ at node $\nu$ of $\tree(H,\pi)$ 
and let $W_i = \{v_1,\dots,v_h\}$ be the target cell of that graph, as determined by a fixed function. Then the tree
rooted at $\nu$ has the trees $\tree(H,\indiA{\pi_\nu}{v_1}),\dots,\tree(H,\indiA{\pi_\nu}{v_h})$ as children.
\end{compactitem}

We observe that: (i) isomorphic graphs have isomorphic search trees, due to the fact that both
the target cell selector and the refinement function are isomorphism invariant; (ii) 
isomorphic leaf nodes of $\tree(G)$ allow to detect automorphisms of $G$ (see e.g.\ \cite{BabaiHand, kocay,McKay81}). 

The typical behavior of algorithms based on the individualization-refinement
mechanism is exemplified by the backtrack search in Figure \ref{f:searchtree}, where each node
of the tree is labeled by an individualized vertex, and is connected by a dashed line to the
graph represented by that node.
Only part of the whole backtrack tree is actually generated. The other
parts of the tree are either shown to be equivalent to parts already generated, or are pruned
by means of invariant information discovered while traversing the tree itself.

Once it has been computed, the leftmost leaf of the tree is stored for comparing 
it with subsequent leaves, in order to find automorphisms of the graph. Such an
automorphism is found when two discrete partitions induce the same quotient. 
In this case the backtrack search restarts from the next individualized vertex
of the least common ancestor of the two leaves which determine the automorphism.
This mechanism guarantees, by the orbit-stabilizer theorem (see e.g.\ \cite{bigwhi}), the correctness of the algorithm
for computing generators and size of the automorphism group of the input graph.

In order to compute the canonical labeling a further leaf is stored, which is the best one according
to some initially defined ordering, and is updated if necessary during the traversal of the tree.

For example, in Figure \ref{f:searchtree}, $Q(H,\refind{4,6})$ is equal to $Q(H,\refind{2,6})$. 
The corresponding generator for the automorphism group of $G$ is $\gamma=(1,9)(2,4)(5,10)$. 
It follows from the existence of an automorphism of $G$ which maps vertex $2$ onto vertex $4$ that the whole 
tree rooted at $2$ carries the same information as the one rooted at $4$; therefore the computation 
of the latter can be cancelled without loss of information.

Instead, looking at the next backtrack step we observe that $Q(H,\refind{6})$ is different from $Q(H,\refind{2})$. 
This implies the non-existence of any automorphism mapping vertex $2$ onto vertex $6$.
In the example of Figure \ref{f:searchtree} we have pruned the tree at the current node, since it has been
assumed that it will not produce a ``better'' canonical form than the initially stored one (which in this
case coincides with the leftmost one). 

The algorithm terminates producing a set of generators for the automorphism group of the input graph $G$,
and the canonical form of $G$.

\section{Introducing \emph{Traces}}\label{s:mot}

The main tools existing in the literature for canonical labeling of graphs and/or automorphism group computation
by means of the individualization-refinement technique are \emph{nauty} \cite{McKay81, nauty22}, \emph{saucy} 
\cite{DargaLSM04,saucy} (recently improved in \cite{saucy2}) and \emph{bliss} \cite{JunttilaK07,bliss}. 
Other tools, such as Kocay's \emph{Groups\&Graphs} \cite{kocGG}, Leon's \emph{partition backtrack} algorithms \cite{leon,leonPB}
and Kreher and Stinson's \cite{KraherStinson1999} software, being of a more general nature, will not be considered for comparison here.

We mention some relevant differences among the three selected tools:
\begin{compactitem}
\item \emph{nauty} implements some invariants allowing finer partitions from the refinement process, in order to speed-up the computation for suitable classes of graphs;
\item \emph{saucy} is especially tailored for sparse graphs, and it computes the automorphism group for the input graph, only,
without considering the canonical labeling problem;
\item \emph{bliss} uses efficient data structures and new heuristics for computing partitions and refinements, and for traversing the search space.
\end{compactitem}
It is important to observe that, if we disregard implementation details, each of these tools is based on the same 
conception, the one described in the previous section. There are four main elements in this
algorithmic structure:
\begin{compactitem}
\item the strategy for building and pruning the search tree, which is the depth-first one;
\item the absence of specific tools for manipulating information coming from the group of detected automorphisms;
\item the refinement procedure, known as $1$-dimensional Weisfeiler-Lehman algorithm \cite{wl} or \emph{vertex classification};
\item the target cell selector, based on local properties of adjacencies of nodes.
\end{compactitem}
The tool we are going to introduce in the present paper is named \emph{Traces}: it is based on different designs
for dealing with all four of the aforementioned issues.

\subsection{Search space construction.}
The central innovation of \emph{Traces} over other canonical labeling algorithms
concerns the search space construction strategy.

We observe that a backtrack search (i.e.\ a depth-first visit of the search space) may cause 
inefficiencies when, at some level, it is possible to \emph{prune the tree by a node invariant}.
This happens when two nodes appearing at the same level in $\tree(H,\pi)$ are associated to partitions which induce different
quotient graphs. Therefore, one of the subtrees rooted at those nodes could be pruned.
In a case such as this, a depth-first strategy might force visiting a whole subtree which will later be discarded.

\begin{figure}[t]
\centering
\includegraphics[scale=.22]{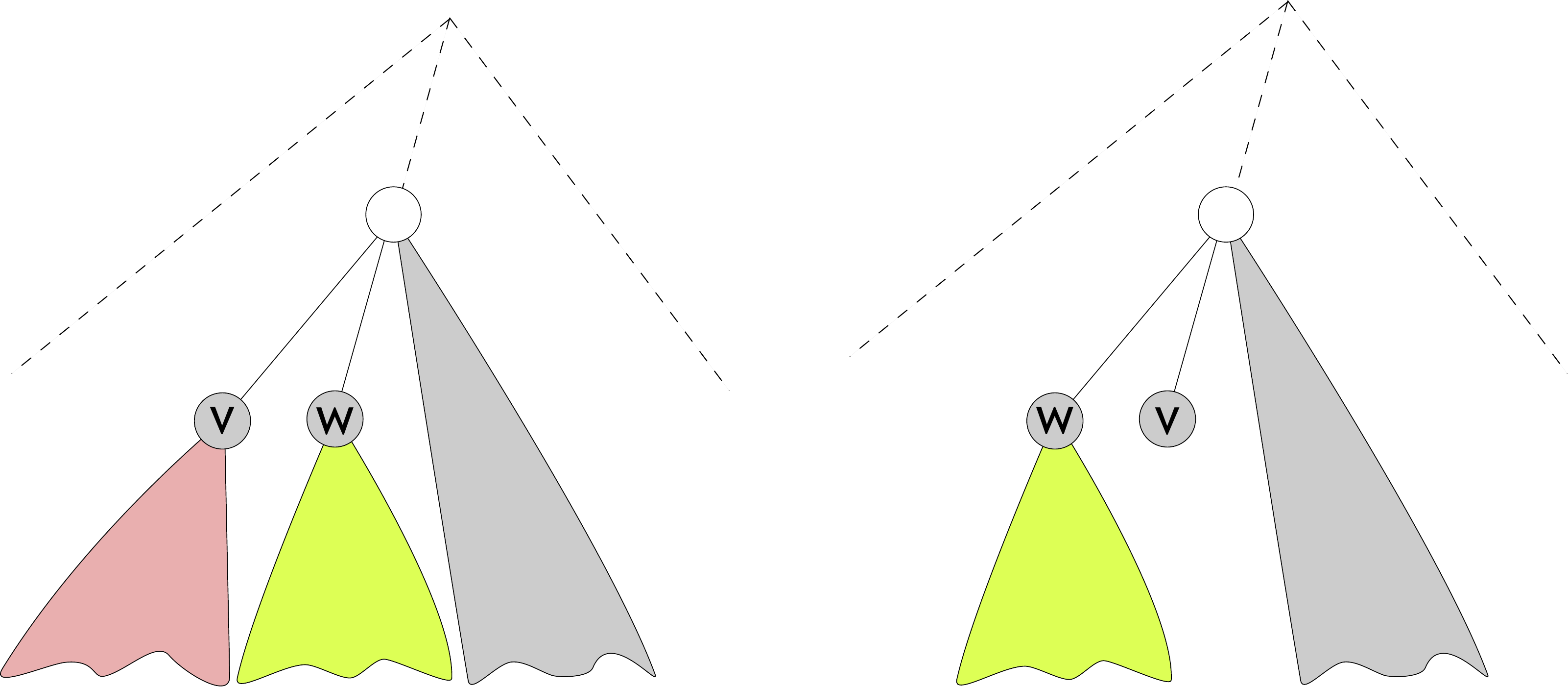}
\caption{Inefficiency of the backtrack strategy}
\label{f:backtrack}
\end{figure}

As an example, let us assume, as depicted in Figure \ref{f:backtrack}, that the quotient graph associated to the node
labeled by $w$, $Q(H,\refind{\vec{u},w})$ is ``better'', according to some predefined ordering, e.g.\ lexicographic, 
than $Q(H,\refind{\vec{u},v})$. In other words, we are assuming that the canonical form is not associated to a leaf of the tree
rooted at $v$.
If $v$ comes before $w$ during the construction of $\tree(H,\pi)$, then the whole subtree rooted at $v$ 
is visited before it is realized that its construction could have been avoided (Figure \ref{f:backtrack}, left). Conversely,
if $w$ comes before $v$ during the construction of $\tree(H,\pi)$, then only the root of that subtree
(which might, in fact, be huge) is visited (Figure \ref{f:backtrack}, right). 
It turns out that the efficiency in pruning the search tree depends strictly on 
the order in which vertices are stored in the target cell, but this order is unpredictable, since cells are sets. 
Furthermore, the circumstances that have just been described are a cause of instability for the whole canonization process: 
completely different performances can be obtained from isomorphic instances of the same graph. 
In a certain sense we can say that
a depth-first search is incapable of capturing the structure of a graph, since its efficiency depends on the
graph's representation.

These considerations suggest it would be better to implement a breadth-first strategy for building the search space, thus 
 enabling all the subtrees whose root does not produce the ``best'' partition to be pruned, for each level of $\tree(H,\pi)$.

However, we must observe that a breadth-first strategy does not allow for
pruning of the search tree by means of automorphism detection, because automorphisms are 
discovered by comparing the leaves of the tree.  Therefore, a simple breadth-first strategy 
would only be able to detect automorphisms during its final iteration. 

Our aim is to define a strategy which combines the advantages of a breadth-first traversal
(early pruning of useless subtrees) with those of a depth-first search (automorphism detection).
Consequently we propose to use the following variant of the breadth-first strategy for traversing the search space:

\begin{quote}
\emph{ for any level $\ell$ of $\tree(H,\pi)$ and for any node $\nu$ appearing at $\ell$, either $\nu$ is discarded or one and only one
path toward a leaf of $\tree(H,\pi)$ is computed.  }
\end{quote}

This path will be called an \emph{experimental path}.
In particular, we will have: (i) for any level $\ell$, non-discarded nodes at $\ell$ share the same
quotient graph; (ii) the computation of the experimental path is started only if $\nu$ is not
known to be equivalent (by automorphism) to some previously computed node at level $\ell$.

\begin{figure}[htbp]
\centering
\includegraphics[scale=.168]{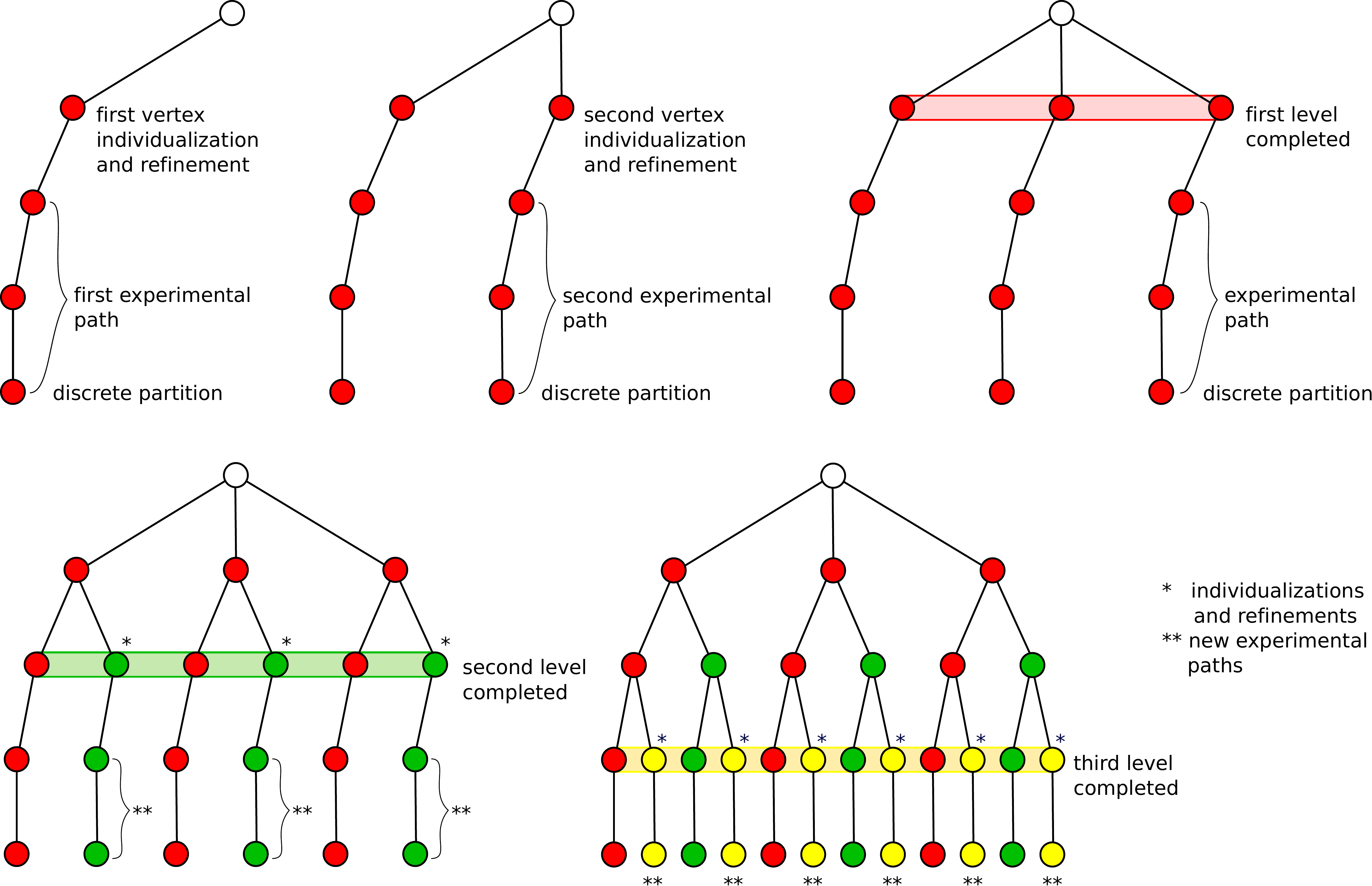}
\put(-176.86,131.6){\scriptsize{$\pi_2$}}
\put(-309.6,131.6){\scriptsize{$\pi_1$}}
\put(-316,-8){\scriptsize{(d)}}
\put(-150,-8){\scriptsize{(e)}}
\put(-90,120){\scriptsize{(c)}}
\put(-250,120){\scriptsize{(b)}}
\put(-340,120){\scriptsize{(a)}}
\caption{Traversing the search space with \emph{Traces}.}
\label{f:vis}
\end{figure}
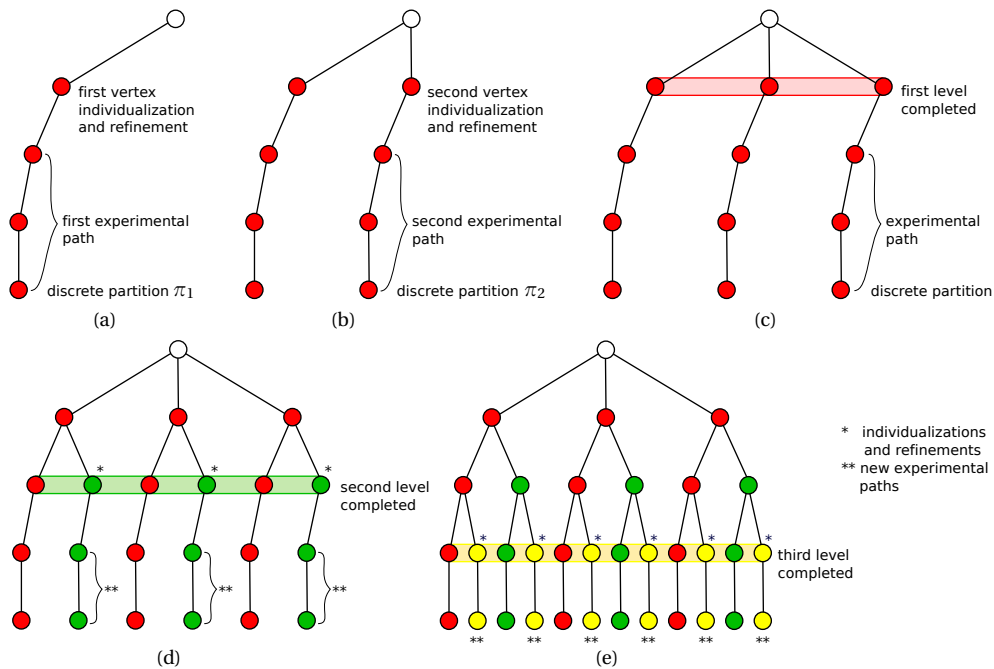

Figure \ref{f:vis} shows the behavior of the search strategy we have just defined. In order to focus on the
traversal we have omitted the pruning operations, which will be introduced in the
definition of the canonical labeling algorithm. In Figure \ref{f:vis}.(a), after
the individualization-refinement of the first vertex in the target cell, an experimental path is
shown, ending with the discrete partition $\pi_1$. In (b), the same operation is illustrated for
the second vertex in the target cell. The new experimental path leads to the partition $\pi_2$, which
can now be compared with $\pi_1$, possibly deriving an automorphism of the input graph.
In the affirmative case, the detected automorphism maps the vertices which have been fixed
to produce $\pi_1$ onto those fixed to produce $\pi_2$. In (c), the computation of the first level
is completed. Figure \ref{f:vis}.(d,e) show the traversal of levels $2$ and $3$. Without entering into
implementation details, we assume that an experimental path already computed at some level
is not computed again at subsequent levels.

\subsection{The use of detected automorphisms.}
When a generator $\gamma$ of the automorphism group of the graph is found by comparing two leaves 
$l_1$ and $l_2$ of $\tree(H,\pi)$, \emph{nauty} prunes the search tree at the level where the deepest 
common ancestor of $l_1$ and $l_2$ appears. 
In addition, it stores some information about fixed points and cycles of $\gamma$ to be used later for the
so called \emph{early pruning} of the search tree, namely the pruning by automorphism
which can be obtained without detecting any further isomorphism. This happens in the example of 
Figure \ref{f:searchtree}, where the rightmost subtree of the root (labeled by $8$) is pruned because of 
the presence of a previously found generator which associates $8$ to $6$.

\begin{figure}[htbp]
\centering
\includegraphics[scale=.20]{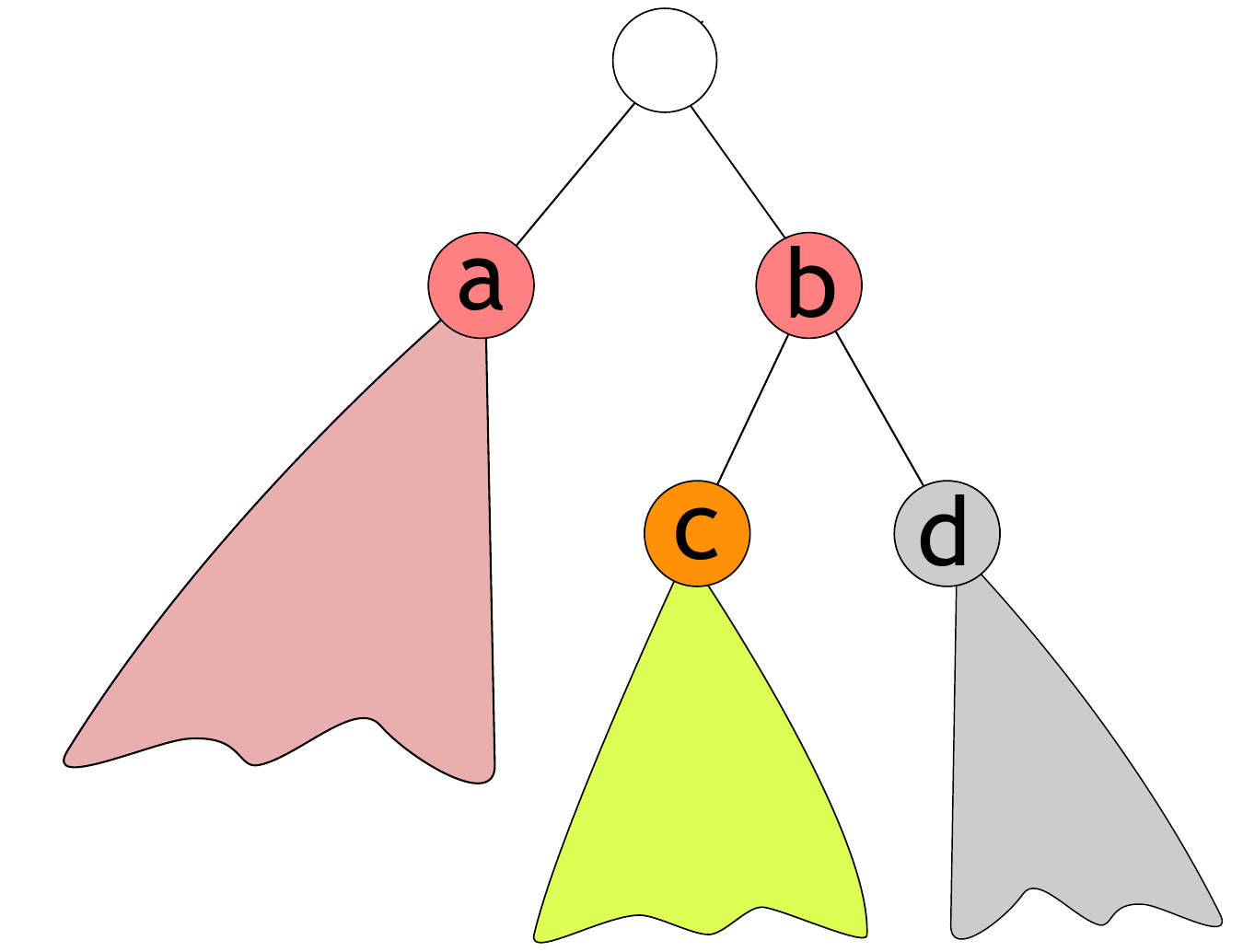}
\caption{A subtree to be pruned by automorphism}
\label{f:prune}
\end{figure}

Such a simple structure, though very powerful in most cases, cannot in general obtain the maximal pruning 
by automorphisms. For example, consider the tree in Figure \ref{f:prune}, where it is assumed that the
exploration of the subtree rooted at $a$ has produced the generators $\gamma_1=\textsf{(v c)(w b)}$ and 
$\gamma_2=\textsf{(v d)(w b)}$ for some $v,w$. The permutation $\gamma=\textsf{(v c d)}$ belongs to the group generated
by $\gamma_1$ and $\gamma_2$. Such a permutation proves that the subtrees rooted at $c$ and $d$ are
equivalent, since it fixes $b$ and sends $c$ to $d$. Therefore, one of the subtrees rooted at $c$ or $d$ should be pruned.
However, it is impossible to establish the equivalence between $c$ and $d$
by looking at $\gamma_1$ and $\gamma_2$ separately, as \emph{nauty} does.

In order to apply pruning by automorphism to its maximal extent, \emph{Traces} uses the Schreier-Sims algorithm
(\cite{sims} , see also \cite{Seress02} for a description and an extensive bibliography on this subject) 
for manipulating the automorphism group of the considered graph. This algorithm
is an efficient method for computing a \emph{base} and a \emph{strong generating set} of a permutation group. It is 
used in \emph{Traces}: (i) for computing the orbits of point stabilizers which correspond to sequences of individualized vertices;
(ii) for computing the size of the automorphism group; (iii) for testing whether an automorphism is already present in the group
generated by the discovered automorphisms; (iv) for avoiding redundant computations during the refinement
process. The Schreier-Sims algorithm is not used by \emph{nauty}, \emph{saucy} and \emph{bliss}.
In each of  these tools only the orbits of stabilizers of vertices which are individualized along the
leftmost path are computed and put to work for pruning the search tree. The Schreier-Sims algorithm
is used by Kreher and Stinson in the canonical labeling algorithm described in \cite{KraherStinson1999}.

\subsection{Refinement procedure and target cell selector.}\label{s:reftc}

A relevant feature of \emph{nauty} is that it gives the user the opportunity of using some sort of invariant
to assist the built-in refinement procedure. The use of invariants enables additional information to be
collected during the refinement process, in order to obtain finer partitions, therefore reducing the
size of the search tree.

Some invariants are very useful for several families of difficult graphs. However, their selection is left by 
\emph{nauty} to the user, as the use of a vertex invariant requires the identification of the input graph, 
thus contradicting the assumption of a general purpose algorithm. 

In the present paper, we experiment the option for the user to adopt 
the $2$-dimensional Weisfeiler-Lehman refinement procedure 
\cite{wl,BabaiHand} as an \emph{universal} invariant, to be applied to any hard graph without consideration of its family.  

\begin{algorithm}[t]
\caption{$2$-dimensional refinement}          
\label{a:refine}                           
\begin{algorithmic}[1]                
\REQUIRE A colored graph $G=(H,\pi)$\vspace{1mm}
\REPEAT 
\STATE Apply $1$-dimensional refinement to $(H,\pi)$, thus updating $\pi$
\FOR{\textbf{each} vertex $v$ in any non-singleton cell of $\pi$}
\STATE Classify $v$ according to the isomorphism type of the quotient of $R(H,\indi{v})$
\ENDFOR 
\STATE Update $\pi$ according to the new classification
\UNTIL no refinement occurred at step $4$
\end{algorithmic}
\end{algorithm}

In the $2$-dimensional Weisfeiler-Lehman refinement, ordered pairs of vertices of a graph $G=((\enne,E),\pi)$
are colored, initially using three colors: edges, non-edges, and the diagonal; in particular, the diagonal coloring reflects
$\pi$. The coloring is iteratively refined by classifying edges according to the number of colored triangles they participate in.
The algorithm stops when no further refinement is possible: the final coloring of diagonal elements is called
a \emph{$2$-stable} partition of vertices of $G$. 
The refinement process is usually represented (see e.g.\  \cite{bast,cam}) by a $n\times n$ matrix
$\textbf{W}$ whose entry $w_{ij}$ is the color of the edge or non-edge $(i,j)$, if $i\neq j$, the color 
of vertex $i$ otherwise ($1\leq i,j\leq n$). We observe that, if $\textbf{W}$ represents a $2$-stable partition, then for
$1\leq i\leq n$ the $i$-th row of $\textbf{W}$ identifies a stable partition $\pi_i$ of vertices of $G$ with singleton
vertex $i$. In addition, if $w_{ii}\neq w_{jj}$ for some $i,j$, then the isomorphism class of the quotient graph
of $G_i=((\enne,E),\pi_i)$ is different from that of $G_j=((\enne,E),\pi_j)$. This suggests the definition of Algorithm \ref{a:refine}
for computing $2$-dimensional refinement.

Let $R$ be the $1$-dimensional Weisfeiler-Lehman refinement function (namely, \emph{nauty}'s refinement). 
In Algorithm \ref{a:refine}, every vertex $v$ of the graph $G=(H,\pi)$ is classified according to the isomorphism type of 
$Q(R(H,\indi{v}))$, the quotient of the graph obtained by $R$ after individualizing $v$. The resulting ranking
yields a new partition of vertices. The classification is repeated until a stable partition is reached.

It must be observed that the $2$-dimensional Weisfeiler-Lehman refinement is computationally much heavier than 
the $1$-dimensional one --- as well as for \emph{nauty}'s invariants, computation is added to the refinement process. 
In the worst case we could be losing a factor of $n$ (the number of vertices of the 
considered graph) in time (\cite{BabaiHand}), since the refinements of all vertex individualizations are needed to stabilize the partition. 
In the next two sections, we introduce some techniques which are implemented in \emph{Traces} in order to simplify the
$2$-dimensional Weisfeiler-Lehman refinement and to obtain some additional information from it.

In fact, a motivation for using the $2$-dimensional algorithm is to collect information from it in order 
to select the target cell to be considered at the next individualization step. In particular,
during the classification of a vertex $v$ (step $3$ of the algorithm),  the number of cells of
$R(H,\indi{v})$ is computed. The target cell associated to the whole refinement process
is chosen as the leftmost one whose vertices produce the maximum of such values.
Experiments reveal that this kind of \emph{look-ahead} technique is useful for decreasing both
the size and the the depth of the search tree.

\begin{remark}
 Two negative results must be mentioned while considering \emph{nauty's} refinement procedure and target cell selector:
(i) Miyazaki's sequence of graphs (\cite{miyazaki}) showing the exponential behavior of \emph{nauty}; (ii)
Cai, F\"urer and Immerman's construction (\cite{CaiFI92}) about (non) identification of graphs via the
generalized $k$-dimensional Weisfeiler-Lehman refinement. 
Miyazaki proved that the choice of the target cell can be responsible for the existence
of intractable graphs for \emph{nauty}. On the other hand, in \cite{CaiFI92} the authors show that there does not exist $k$
such that the $k$-dimensional refinement is able to capture the orbit partition of any graph.

As a matter of fact, while Miyazaki's result has a negative impact on \emph{nauty} and other practical isomorphism tools, 
Cai, F\"urer and Immerman's construction provides a theoretical justification for the individualization-refinement technique, since
it proves that, (at least) in the presence of Weisfeiler-Lehman refinement, it is impossible to detect the orbit partition of the
automorphism group of a graph directly, without the help of a search space construction. 
\end{remark}

\subsection{Comparing partition refinements}
It is crucial for every tool based on the individualization-refinement technique to have
an efficient procedure for computing refinements and for comparing them. 

Refinements are compared in our tool without computing them completely. This possibility has
already been observed by Junttila and Kaski in \cite{JunttilaK07} in the case of singleton
cells emerging during the refinement process (the consequent invariant is called
{\em partial leaf certificate}, and is adopted in \emph{Traces}, too).

In addition to this, we implement an invariant, which we call {\em refinement trace},
based on the following observation: the whole refinement process, namely the sequence of cell splitting, 
is an (isomorphism) invariant, not only its final result. 
Assume that the cell $W$ of the partition $\pi$ is split during the refinement process into $W'$ and $W''$. 
This splitting gives the partition $\pi'$ such that:
$\pos(W',\pi') = \pos(W,\pi) \mbox{ and } k = \pos(W'',\pi') = \pos(W,\pi)+\mathopen\mid{W'}\mathclose\mid$.
The new position $k$ created by splitting the cell $W$ is a \emph{trace element} of the refinement
process. The refinement trace is the sequence of trace elements successively introduced
during refinement; it is isomorphism invariant and can be stored into an array.
Moreover, let us consider the alternation of individualization and refinement steps which
is needed to compute a discrete partition; the whole process has its own trace,
since the individualization operation consists of a cell splitting, too. The trace has a length of
at most $n$, and each of its elements appears exactly once in it. 
Note that, since they are sequences of integers, traces can be ordered, e.g.\ component-wise.

Assume now that the refinement of $G_1=(H_1,\pi_1)$ has been computed and $\tau=(\tau_1,\dots,\tau_m)$
is its trace. Assume also that, while computing the refinement of $G_2=(H_2,\pi_2)$, 
there exists a trace element $\tau'_i$ which is different from the corresponding $\tau_i$. 
This is sufficient to establish that the refinements of $G_1$ and $G_2$ will be different. 
Obviously, if we choose an ordering on traces and $\tau'_i$ is ``better'' than $\tau_i$,
then the refinement of $G_2$ will be completed and its trace will be stored for later comparisons.

A similar mechanism is implemented in \emph{saucy} \cite{saucy2} for comparing refinements. The main difference
is that \emph{saucy} considers the trace of the leftmost backtrack path, only. All subsequent refinements
are compared against this trace. In our algorithm the trace is updated as soon as a better one is 
found. At the end of the computation the best trace will be obtained, namely the one which is
associated to the canonical form of the input graph.

\subsection{$2$-dimensional Weisfeiler-Lehman refinement and automorphisms}\label{s:2dimaut}
The $2$-dimensional refinement procedure (Algorithm \ref{a:refine}) is used in \emph{Traces} in conjunction with
the Schreier-Sims algorithm, in two different ways.

Given a graph $G=(H,\pi)$, let $\Gamma$ be a subgroup of its automorphism group.
Let us assume that, for some $k\geq0$ and for some vertices $v_1,\dots,v_k$, the graph
$G'=(H,\refind{v_1,\dots,v_k})$ appears at a node of $\tree{(G)}$. Let $w$ be 
a vertex in the target cell associated to $G'$, so that the refinement of 
$(H,\indiA{\refind{v_1,\dots,v_k}}{w})$ must be computed.
Using the Schreier-Sims algorithm, the orbits of the stabilizer of $\Gamma$ with
respect to $v_1,\dots,v_k,w$ can be computed. This enables the classification step
of Algorithm \ref{a:refine} to be simplified: for each cell of $\indiA{\refind{v_1,\dots,v_k}}{w}$,
only one vertex for each orbit is classified, all the other ones being equivalent to it.

On the other hand, when the classification of vertices from a cell produces discrete partitions,
the corresponding colored graphs are compared for checking whether they induce an
automorphism of the graph. Therefore, automorphism can be detected during the refinement
process, too. 

Furthermore, while refining $(H,\indiA{\refind{v_1,\dots,v_k}}{w})$, 
let $W$ be the leftmost cell such that the classification of its vertices 
produces discrete partitions, and let us assume that $W$ is not a singleton cell. 
All vertices in $W$ give the same quotient graph, otherwise Algorithm \ref{a:refine} would split it. 
Therefore, for any $u,v\in W$ there is an automorphism of $G$ which maps $u$ onto $v$.
Moreover, as defined in Section \ref{s:reftc}, $W$ is the target cell returned by the refinement procedure.
As a consequence, we infer that the current node of the search tree will have one leaf as its unique
child. \emph{Traces}' refinement will therefore return this leaf as the result of the current refinement.

\begin{figure}[htbp]
\centering
\includegraphics[scale=.15]{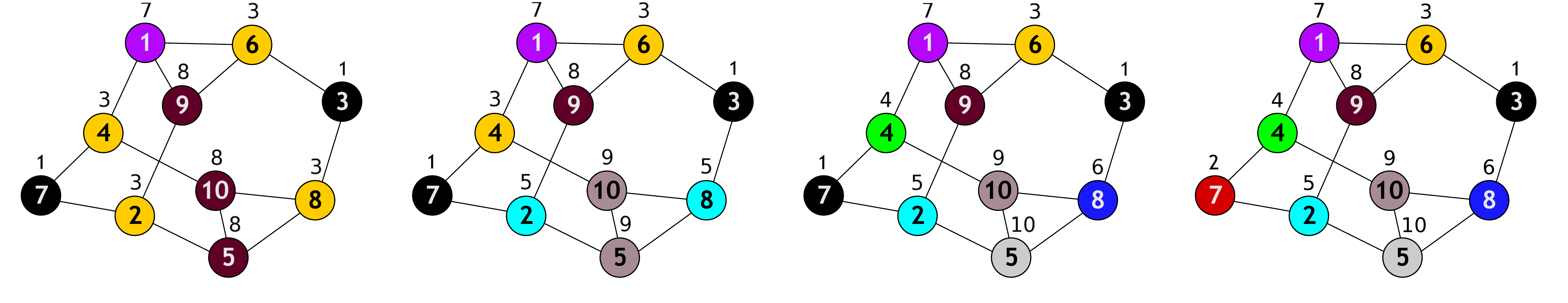}
\caption{(a) Individualization of vertex $1$ and refinement.}
\label{f:G10-5}
\end{figure}

As an example, still considering the graph $G=(H,(\insie{3,7},\insie{2,4,6,8},\insie{1,5,9,10}))$ 
of Figure \ref{f:G10-1} (left), note that $G$ is equitable and hence not refinable by the $1$-dimensional 
Weisfeiler-Lehman algorithm. While applying the $2$-dimensional algorithm, all vertices are classified
by individualizing them and comparing the quotients of the resulting refinements. It turns out that
the third cell produces equivalent discrete partitions, thus allowing for the detection of some
automorphisms of the graph. In particular Figure \ref{f:G10-5} shows that the refinement starting from
the individualization of $1$ yields the partition 
$(\insie{3},\insie{7},\insie{6},\insie{4},\insie{2},\insie{8},\insie{1},\insie{9},\insie{10},\insie{5})$.
Similarly, the refinement starting from the individualization of $5$ yields the partition 
$(\insie{3},\insie{7},$ $\insie{8},\insie{2},\insie{4},\insie{6},\insie{5},\insie{10},\insie{9},\insie{1})$.
Comparison of the partitions thus obtained enables the automorphism \textsf{(1 5)(2 4)(6 8)(9 10)}
to be found.

\section{The Canonical labeling Algorithm}\label{s:canlabalg}
We introduce two orderings on graphs to be used in the description of the canonical labeling algorithm
and in the proof of its correctness.
\begin{definition} Let $\pi_1$ and $\pi_2$ be two stable partitions of vertices of the graph $H$.

(i) We say that the colored graph $G_1 = (H,\pi_1)$ is  \emph{$1$-dim better} than $G_2 = (H,\pi_2)$ 
when either $\pi_1$ has more cells than $\pi_2$, or $\pi_1$ and $\pi_2$ have the same number of cells and $Q(G_1)$ 
is lexicographically smaller than (or equal to) $Q(G_2)$.

(ii) We say that $G_1$ is \emph{$2$-dim better} than $G_2$ when either $G_1$ is $1$-dim better than
$G_2$ or $Q(G_1) = Q(G_2)$ and $Q(H,\refpiind{\pi_1}{v_1})$ is $1$-dim better  than
$Q(H,\refpiind{\pi_2}{v_2})$, where $v_1$ and $v_2$ are elements of the target cell of 
$G_1$ and $G_2$, respectively.

(iii) For $k=1,2$, we say that $G_1$ is  \emph{$k$-dim equal} to $G_2$ if 
$G_1$ is $k$-dim better than $G_2$ and $G_2$ is $k$-dim better than $G_1$.
\end{definition}

Two versions of \emph{Traces}' canonical labeling algorithm are defined in Algorithm \ref{a:canlab},
depending on the use of $1$-dimensional or $2$-dimensional Weisfeiler-Lehman
refinement;  each version exhibits its own target cell selector:
\begin{compactitem}
\item in the case of $1$-dimensional refinement, the target cell will be chosen according to the following rule: 
at the initial level it is the largest one; at level $\ell>0$,
the largest cell which is contained in the target cell at level $\ell-1$ (or at level $\ell-2$ if the target cell at level 
$\ell-1$ has been transformed into singletons, and so on backwards).
\item in the case of $2$-dimensional refinement, the target cell will be selected as described in Section \ref{s:reftc}.
\end{compactitem}

\begin{algorithm}[t]
\caption{Canonical Labeling with $k$-dimensional refinement ($k\in\{1,2\}$)}          
\label{a:canlab}                           
Remark: throughout the algorithm ``the group'' refers to the group generated 
by the successively detected automorphisms;

$^{(\star)}$ use the $k$-dimensional algorithm; $^{(\star \star)}$ in the $2$-dim case, only.
\begin{algorithmic}[1]                
\REQUIRE A colored graph $G$\vspace{1mm}
\STATE Initialize the automorphism group of $G$
\STATE Refine$^{(\star)}$ $G$ to $(H,\pi)$, possibly$^{(\star \star)}$ 
adding new generators to the group
\STATE $\ell=0$ \COMMENT {initial level}
\STATE Build the list of graphs at level $0$, whose element is $(H,\pi)$, only
\WHILE{the number of cells of partitions at the current level $\ell$ is less than $n$} 
\WHILE {the list of graphs at the current level is not empty}
\STATE Consider the head of the list, which, for some $v_{h_1},\dots,v_{h_\ell}$, contains the graph

$(H,\pi^{(v_{h_1},\dots,v_{h_\ell})})$, and remove it from the structure \COMMENT{the current graph}
\FOR{\textbf{each} vertex $v_{h_{\ell+1}}$ in the target cell of the current graph}
\IF{for $i=0,\dots,\ell: v_{h_{i+1}}$ is an orbit representative of the stabilizer of the group 
with respect to $v_{h_1},\dots,v_{h_i}$}
\STATE Produce $(H,\pi^{(v_{h_1},\dots,v_{h_\ell},v_{h_{\ell+1}})})$ after individualization of $v_{h_{\ell+1}}$ 
and refinement$^{(\star)}$, possibly $^{(\star \star)}$
adding new generators to the group
\IF{the obtained graph is $k$-dim better or equal than the best one computed}
\IF{it is $k$-dim better}
\STATE Initialize (as empty) the list of graphs at the next level
\ENDIF
\STATE Compute an experimental path\COMMENT{trivial when the current partition is discrete}
\STATE Check for automorphisms and possibly add new generators to the group
\IF {the next level list is empty \textbf{or} no automorphism is found}
\STATE Append the current graph to the next level list
\ENDIF
\ENDIF
\ENDIF
\ENDFOR \COMMENT{end of vertex individualizations for the current graph}
\ENDWHILE \COMMENT{end of graphs at the current level}
\STATE $\ell=\ell+1$
\STATE Let the list of graphs at the next level become the list at the current level
\ENDWHILE
\end{algorithmic}
\end{algorithm}

\begin{remark}\label{r:canlab}
Note that: (i) the outermost \textbf{while}-loop is executed until a discrete partition is found.
(ii) For each level, the list of graphs to be considered at the next level is built.
(iii) New generators of the automorphism group of the input graph can be found either during the 
refinement process (but only with $2$-dimensional refinement, see Section \ref{s:2dimaut}), or comparing graphs 
coming from experimental paths. 
(iv) In the latter case, the current graph is not added to the list of
graphs to be considered at the next level (see the \textbf{if}-statement at line $17$), since it is equivalent to
a previously computed graph.
(v) Pruning by automorphism is allowed by the \textbf{if}-statement at line $9$, where the orbits of the stabilizer
of the group (with respect to the vertices which are fixed along the path from the root of the search tree to the
current node) are computed by the Schreier-Sims algorithm. For each orbit, only one vertex (a representative)
is individualized. More precisely, for each orbit \emph{Traces} chooses as representative its smallest vertex.
\end{remark}

The following hold with $k=1,2$.
\begin{proposition}\label{p:equal}
For each $\ell$, all the graphs computed by Algorithm \ref{a:canlab} at level $\ell$ are $k$-dim equal.
\end{proposition}
\begin{proof}
Induction on $\ell$. Easy when $\ell=0$, since there is only one graph at the initial level.
During the construction of the list of graphs at the next level, if a better graph is obtained, then the whole list is
re-initialized (line $13$) and all the previously computed graphs are discarded; if a worse graph is obtained,
it is not added to the list (see the \textbf{if}-statements at lines $11$-$12$).
\end{proof}

\begin{lemma}\label{l:lemma1}
For any colored graph $G$, the final level of $\tree(G)$ produced by Algorithm \ref{a:canlab} consists of only one node.
\end{lemma}
\begin{proof}
Follows from Proposition \ref{p:equal}, using the argument of Remark \ref{r:canlab}.(iv).
\end{proof}

\begin{lemma}\label{l:lemma2}
Let $k\in\{1,2\}$. The $k$-dimensional refinement function and the corresponding target cell selector
used by Algorithm  \ref{a:canlab} are invariant under isomorphism.
\end{lemma}
\begin{proof}
It is well known that $1$- and $2$-dimensional Weisfeiler-Lehman refinements are isomorphism
invariant \cite{wl,BabaiHand}. When $k=2$, \emph{Traces} uses the $2$-dimensional refinement at every level of the search tree
except at the final one, where (see Section \ref{s:2dimaut}) a further $1$-dimensional refinement step is executed.
Therefore, the adopted refinement is isomorphism invariant. 

Let us first consider Algorithm  \ref{a:canlab} in the case $k=1$.  
Let $G=(H,\pi)$ and let $S=(a_0=0,b_0=n),(a_1,b_1),\dots,(a_h,b_h)$ be a sequence of pairs on integers such that:
$\forall i\in\{1,\dots,h\}$: (i) $a_i < b_i$; (ii) $a_i$ is the position of a cell of $\pi$; (iii) either $b_i=n$ or 
$b_i$ is the position of a cell of $\pi$. 
Let $\widetilde{h}$ be the largest index in $\{1,\dots,h\}$ such that there exists a non-trivial
cell of $\pi$ whose position $p$ is such that $p\geq a_{\widetilde{h}}$ and $p<b_{\widetilde{h}}$.
The target cell of $G$ is the leftmost cell with maximal size whose position lies within
the interval $[a_{\widetilde{h}},a_{\widetilde{h}})$. 
If $G_1=(H,\pi_1)$ and $G_2=(H,\pi_2)$ are isomorphic and we choose their target cell according to the same sequence
$S$, then the same target cell will be selected for $G_1$ and $G_2$, since the sequences of sizes of cells
in $\pi_1$ and $\pi_2$ are equal. 
It follows by induction on the depth of the search tree, using Proposition \ref{p:equal}, that the sequences
$$S_0=(0,n); S_{\ell+1}=S_\ell,(a,b)$$ where $a$ is the position of the target cell $W_a$ at level $\ell$ 
according to $S_\ell$ and $b=a+|W_a|$, are isomorphism invariant, as well as the target cell selector.

In the $2$-dimensional case the adopted target cell selector is trivially isomorphism invariant by construction.
\end{proof}

\begin{proposition}[Correctness]
Given two colored graphs $G_1$ and $G_2$, let $Q_1$ and $Q_2$ be the quotient graphs
associated to the graphs produced by Algorithm \ref{a:canlab} at its final iteration, respectively. 
If $G_1$ and $G_2$ are isomorphic, then $Q_1=Q_2$.
\end{proposition}
\begin{proof}
From Proposition \ref{p:equal} and from Lemma \ref{l:lemma2}, we have that for any $\ell\geq0$ and
for any graph appearing at level $\ell$ in $\tree(G_1)$ there exists one graph isomorphic to it at level $\ell$ in 
$\tree(G_2)$. The correctness of Algorithm \ref{a:canlab} follows from Lemma \ref{l:lemma1}.
\end{proof}

\section{Experimental results}\label{s:exper}

The algorithm presented in the paper is now compared with \emph{nauty} and \emph{bliss}.  A comparison with \emph{saucy} 
can be obtained from our performance tables and those presented in \cite{JunttilaK07}. 
We observe that \emph{saucy} is extremely efficient on some families of graphs which
mainly come from encodings of the satisfiability problem, while it is
usually slower than all the other tools outside those classes. Moreover, 
\emph{saucy} does not compute a canonical labeling.

\subsection{Methodological statement}
\emph{Traces} implements a general purpose algorithm aimed at reducing the search space associated to canonical labeling
and/or to the computation of the automorphism group of a graph. The algorithm which only computes a set of generators
for the automorphism group of the input graph - fully described in \cite{prelim} - is obtained by means of a slight 
modification of the canonical labeling algorithm. 

Every canonical labeling tool based on the individualization-refinement technique spends most of its execution time in
refining partitions. We take $1$-dimensional refinement as the unit of measurement of the size of the search
tree, since one refinement is completed for each node of the tree by all canonical labeling tools. 
In the case of experiments with $2$-dimensional refinement, the number of $1$-dimensional steps 
needed during the refinement process will be counted and reported in tables.

The selected benchmarks are divided in two parts: (i) those displaying a large search space (hard graphs), and (ii)
those with a small search space (easy graphs). The use of $2$-dimensional refinement is not considered for easy graphs.

In our experiments, whenever possible, we have chosen the appropriate invariant to establish the best performance
for \emph{nauty}. It has to be noted that the choice of a correct invariant is a subtle operation; as an example,
the invariant \emph{cellfano2} is useful to break the regularity of incidence graphs of projective planes, as it
looks for occurrences of Fano subplanes into the considered graph. But it reveals itself to be extremely
inefficient in the case of projective planes with very large automorphism group (e.g.\ \textsf{pp16-1}). 

\begin{remark}
At present, \emph{Traces} is implemented as an additional command of \emph{nauty}: it uses the
data structures of \emph{nauty 2.4}, with the exception of the graph representation 
(adjacency matrix in \emph{nauty}, adjacency list in \emph{Traces}). 
Indeed, the current version of \emph{Traces} is a prototype which has been implemented with
the intention of substantiating from the experimental standpoint the algorithmic design described in
this paper, with a particular attention to the new search strategy and the use of the Schreier-Sims algorithm. 
In particular, \emph{Traces} does not implement any special data structure or procedure to handle large 
sparse graphs. 

Leon's implementation of the Schreier-Sims algorithm is used in the present version of
\emph{Traces} for automorphism group computations. This code is part of a more general package (\cite{leon}) and
we expect a substantial improvement from a new implementation of the Schreier-Sims algorithm, specialized
for our purposes, which is currently under development.
\end{remark}

\renewcommand\arraystretch{1}
\renewcommand{\tabcolsep}{1pt}
\begin{table}[t]
\begin{scriptsize}
\begin {tabular}{|c|c|r|r|c|c||c|r<{\pgfplotstableresetcolortbloverhangright }@{}l<{\pgfplotstableresetcolortbloverhangleft }|r<{\pgfplotstableresetcolortbloverhangright }@{}l<{\pgfplotstableresetcolortbloverhangleft }||r<{\pgfplotstableresetcolortbloverhangright }@{}l<{\pgfplotstableresetcolortbloverhangleft }|r<{\pgfplotstableresetcolortbloverhangright }@{}l<{\pgfplotstableresetcolortbloverhangleft }||r<{\pgfplotstableresetcolortbloverhangright }@{}l<{\pgfplotstableresetcolortbloverhangleft }|r<{\pgfplotstableresetcolortbloverhangright }@{}l<{\pgfplotstableresetcolortbloverhangleft }|}%
\hline  & & & & \multicolumn {1}{c|}{ }& &\multicolumn {5}{c||}{{\bf\emph{nauty 2.4}}}&\multicolumn {4}{c||}{\bf \emph{bliss 0.50}}&\multicolumn {4}{c|}{\bf\emph{Traces}}\\%
Graph&Ref&V&E&$\mid \text {Aut}\mid $&Orbs&\multicolumn {1}{c}{Inv}&\multicolumn {2}{c}{Time}&\multicolumn {2}{c||}{Size}&\multicolumn {2}{c}{Time}&\multicolumn {2}{c||}{Size} &\multicolumn {2}{c}{Time}&\multicolumn {2}{c|}{Size}\\\hline %
pp16-1&\cite {bliss}&546&4641&$>3.42\cdot 10^{10}$&1&--&$0$&$.05$&$127$&$$&$0$&$.01$&$100$&$$&$0$&$.07$&$841$&$$\\%
pp16-2&\cite {bliss}&546&4641&$2^{8}{\per }3{\per }5$&10&CF&$62$&$.90$&$10$&$$&$671$&$.53$&$46{,}005{,}059$&$$&$35$&$.28$&$110{,}215$&$$\\%
pp16-4&\cite {bliss}&546&4641&$2^{12}{\per }3$&6&CF&$74$&$.22$&$23{,}986$&$$&$104$&$.33$&$8{,}164{,}407$&$$&$12$&$.53$&$38{,}054$&$$\\%
pp16-6&\cite {bliss}&546&4641&$2^{11}{\per }3^{2}$&5&CF&$61$&$.78$&$12$&$$&$3{,}152$&$.94$&$539{,}781{,}990$&$$&$15$&$.53$&$55{,}631$&$$\\%
pp16-7&\cite {bliss}&546&4641&$2^{14}{\per }3^{2}$&3&CF&$61$&$.61$&$4979$&$$&$576$&$.89$&$81{,}992{,}440$&$$&$1$&$.22$&$6{,}323$&$$\\%
pp16-8&\cite {bliss}&546&4641&$2^{15}{\per }3^{3}$&3&CF&$62$&$.55$&$727$&$$&$1$&$.88$&$199{,}505$&$$&$0$&$.36$&$4{,}561$&$$\\%
pp16-9&\cite {bliss}&546&4641&$2^{12}{\per }3^{2}{\per }5^{2}$&6&CF&$60$&$.96$&$39$&$$&$200$&$.69$&$18{,}774{,}117$&$$&$0$&$.36$&$7{,}288$&$$\\%
pp16-11&\cite {bliss}&546&4641&$2^{12}{\per }3^{2}{\per }7$&6&CF&$60$&$.27$&$88$&$$&$301$&$.97$&$45{,}204{,}426$&$$&$0$&$.89$&$16{,}971$&$$\\%
pp16-15&\cite {bliss}&546&4641&$2^{11}{\per }3^{3}$&8&CF&$60$&$.34$&$42$&$$&$1{,}712$&$.23$&$230{,}978{,}343$&$$&$3$&$.47$&$29{,}545$&$$\\%
pp16-17&\cite {bliss}&546&4641&$2^{11}{\per }3^{2}{\per }5$&8&CF&$60$&$.29$&$151$&$$&$66$&$.03$&$5{,}025{,}112$&$$&$1$&$.89$&$31{,}465$&$$\\%
pp16-19&\cite {bliss}&546&4641&$2^{8}{\per }3^{2}$&14&CF&$59$&$.20$&$10$&$$&$839$&$.73$&$63{,}454{,}501$&$$&$61$&$.68$&$183{,}989$&$$\\%
pp16-21&\cite {bliss}&546&4641&$2^{7}{\per }3^{3}$&12&CF&$59$&$.14$&$7$&$$&$4{,}991$&$.69$&$518{,}875{,}221$&$$&$43$&$.11$&$127{,}208$&$$\\%
pp25&\cite {moorhouse}&1302&16926&$2^{7}{\per }3{\per }5^{3}{\per }31$&2&CF&$\textbf{5{,}000}$&$$&$20$&$$&$2{,}456$&$.04$&$118{,}865{,}645$&$$&$13$&$.59$&$6{,}836$&$$\\%
pp27&\cite {moorhouse}&1514&21196&$2^{3}{\per }3^{7}{\per }7$&4&CF&$\textbf{5{,}000}$&$$&$3$&$$&$1{,}704$&$.90$&$116{,}513{,}018$&$$&$431$&$.23$&$109{,}978$&$$\\%
mz-aug2-18&\cite {bliss}&432&684&$2^{38}$&252&--&$66$&$.00$&$5{,}374{,}331$&$$&$7$&$.39$&$1{,}406{,}880$&$$&$1$&$.04$&$73{,}415$&$$\\%
mz-aug2-20&\cite {bliss}&480&760&$2^{42}$&280&--&$337$&$.76$&$23{,}573{,}421$&$$&$30$&$.11$&$5{,}639{,}006$&$$&$1$&$.25$&$77{,}379$&$$\\%
mz-aug2-22&\cite {bliss}&528&836&$2^{46}$&308&--&$1{,}616$&$.68$&$102{,}760{,}999$&$$&$123$&$.07$&$22{,}587{,}583$&$$&$1$&$.54$&$94{,}211$&$$\\%
mz-aug2-30&\cite {bliss}&720&1140&$2^{62}$&420&--&$\textbf{5{,}000}$&$$&$225{,}143{,}802$&$$&$\textbf{5{,}000}$&$$&$797{,}139{,}792$&$$&$3$&$.42$&$178{,}099$&$$\\%
mz-aug2-50&\cite {bliss}&1200&1900&$2^{102}$&700&--&$\textbf{5{,}000}$&$$&$140{,}555{,}243$&$$&$\textbf{5{,}000}$&$$&$832{,}923{,}145$&$$&$17$&$.57$&$597{,}191$&$$\\%
had-52&\cite {bliss}&208&5512&$2^{4}{\per }13$&2&CQ&$0$&$.09$&$13$&$$&$0$&$.29$&$13{,}082$&$$&$0$&$.20$&$2{,}678$&$$\\%
had-100&\cite {bliss}&400&20200&$2^{4}{\per }5^{2}$&2&CQ&$1$&$.62$&$13$&$$&$2$&$.62$&$53{,}932$&$$&$2$&$.69$&$6{,}503$&$$\\%
had-184&\cite {bliss}&736&68080&$2^{6}{\per }23$&2&CQ&$33$&$.40$&$107$&$$&$22$&$.39$&$115{,}937$&$$&$9$&$.86$&$16{,}633$&$$\\%
had-232&\cite {bliss}&928&108112&$2^{6}{\per }29$&2&CQ&$102$&$.14$&$128$&$$&$50$&$.81$&$181{,}668$&$$&$25$&$.33$&$19{,}307$&$$\\%
had-sw-32-1&\cite {bliss}&128&2112&$2^{2}$&42&CQ&$0$&$.06$&$66$&$$&$2$&$.49$&$124{,}818$&$$&$0$&$.55$&$18{,}584$&$$\\%
had-sw-88&\cite {bliss}&352&15664&$2^{2}$&132&CQ&$20$&$.32$&$140$&$$&$204$&$.90$&$3{,}147{,}799$&$$&$35$&$.47$&$160{,}480$&$$\\%
had-sw-112&\cite {bliss}&448&25312&$2^{2}$&224&CQ&$99$&$.63$&$223$&$$&$560$&$.90$&$10{,}990{,}338$&$$&$198$&$.76$&$402{,}977$&$$\\%
sts-67&\cite {bliss}&737&35376&$3$&253&AT&$0$&$.12$&$3$&$$&$8$&$.31$&$157{,}566$&$$&$1$&$.63$&$510$&$$\\%
sts-sw-21-10&\cite {bliss}&70&945&$1$&70&AT&$0$&$.01$&$1$&$$&$0$&$.02$&$3{,}011$&$$&$0$&$.01$&$142$&$$\\%
sts-sw-55-1&\cite {bliss}&495&19305&$1$&495&AT&$0$&$.04$&$1$&$$&$7$&$.58$&$206{,}416$&$$&$1$&$.19$&$992$&$$\\%
sts-sw-79-11&\cite {bliss}&1027&58539&$1$&1027&AT&$0$&$.24$&$1$&$$&$69$&$.79$&$937{,}652$&$$&$14$&$.36$&$2056$&$$\\%
GenQuad-1&\cite {royleprivate}&2752&481600&$2^{8}{\per }3^{2}{\per }7^{5}$&2&--&$\textbf{5{,}000}$&$$&$3{,}802{,}052$&$$&$981$&$.98$&$1{,}291{,}025$&$$&$2$&$.60$&$17857$&$$\\%
Hypercube&\cite {mckayprivate}&3161&18780&$2^{5}{\per }3^{2}$&42&--&$1{,}215$&$.88$&$186{,}798$&$$&$32$&$.45$&$146{,}987$&$$&$2$&$.59$&$49961$&$$\\\hline %
\end {tabular}
\end{scriptsize}
\vspace*{1.6mm}
\caption{Graphs with large search tree: Canonical Form 
(\emph{nauty} invariants \cite{nauty22}: CF = cellfano2, CQ~=~cellquads, AT = adjtriang)}
\label{t:res_can}
\end{table}

\renewcommand\arraystretch{1}
\renewcommand{\tabcolsep}{1pt}
\begin{table}[t]
 \fontsize{7.4}{9}
 \selectfont
\begin {tabular}{|c|c|r|r|c|c||c|r<{\pgfplotstableresetcolortbloverhangright }@{}l<{\pgfplotstableresetcolortbloverhangleft }|r<{\pgfplotstableresetcolortbloverhangright }@{}l<{\pgfplotstableresetcolortbloverhangleft }||r<{\pgfplotstableresetcolortbloverhangright }@{}l<{\pgfplotstableresetcolortbloverhangleft }|r<{\pgfplotstableresetcolortbloverhangright }@{}l<{\pgfplotstableresetcolortbloverhangleft }||r<{\pgfplotstableresetcolortbloverhangright }@{}l<{\pgfplotstableresetcolortbloverhangleft }|r<{\pgfplotstableresetcolortbloverhangright }@{}l<{\pgfplotstableresetcolortbloverhangleft }|}%
\hline  & & & & \multicolumn {1}{c|}{ }& &\multicolumn {5}{c||}{{\bf\emph{nauty 2.4}}}&\multicolumn {4}{c||}{\bf \emph{bliss 0.50}}&\multicolumn {4}{c|}{\bf\emph{Traces}}\\%
Graph&Ref&V&E&$\mid \text {Aut}\mid $&Orbs&\multicolumn {1}{c}{Inv}&\multicolumn {2}{c}{Time}&\multicolumn {2}{c||}{Size}&\multicolumn {2}{c}{Time}&\multicolumn {2}{c||}{Size} &\multicolumn {2}{c}{Time}&\multicolumn {2}{c|}{Size}\\\hline %
pp16-1&\cite {bliss}&546&4641&$>3.42\cdot 10^{10}$&1&--&$0$&$.05$&$144$&$$&$0$&$.01$&$144$&$$&$0$&$.08$&$841$&$$\\%
pp16-2&\cite {bliss}&546&4641&$2^{8}{\per }3{\per }5$&10&CF&$60$&$.37$&$10$&$$&$976$&$.65$&$80{,}597{,}650$&$$&$2$&$.15$&$22{,}194$&$$\\%
pp16-4&\cite {bliss}&546&4641&$2^{12}{\per }3$&6&CF&$70$&$.89$&$27857$&$$&$115$&$.02$&$10{,}311{,}534$&$$&$2$&$.23$&$8{,}569$&$$\\%
pp16-6&\cite {bliss}&546&4641&$2^{11}{\per }3^{2}$&5&CF&$61$&$.97$&$12$&$$&$5{,}386$&$.10$&$968{,}486{,}421$&$$&$0$&$.83$&$28{,}561$&$$\\%
pp16-7&\cite {bliss}&546&4641&$2^{14}{\per }3^{2}$&3&CF&$66$&$.30$&$17859$&$$&$1{,}445$&$.07$&$210{,}159{,}039$&$$&$0$&$.35$&$3{,}256$&$$\\%
pp16-8&\cite {bliss}&546&4641&$2^{15}{\per }3^{3}$&3&CF&$62$&$.18$&$857$&$$&$10$&$.80$&$1{,}170{,}686$&$$&$0$&$.20$&$5{,}126$&$$\\%
pp16-9&\cite {bliss}&546&4641&$2^{12}{\per }3^{2}{\per }5^{2}$&6&CF&$62$&$.47$&$183$&$$&$353$&$.70$&$35{,}107{,}649$&$$&$0$&$.23$&$9{,}290$&$$\\%
pp16-11&\cite {bliss}&546&4641&$2^{12}{\per }3^{2}{\per }7$&6&CF&$61$&$.79$&$56$&$$&$1{,}388$&$.90$&$219{,}395{,}299$&$$&$0$&$.27$&$25{,}514$&$$\\%
pp16-15&\cite {bliss}&546&4641&$2^{11}{\per }3^{3}$&8&CF&$61$&$.75$&$135$&$$&$2{,}003$&$.88$&$278{,}419{,}116$&$$&$0$&$.56$&$29{,}566$&$$\\%
pp16-17&\cite {bliss}&546&4641&$2^{11}{\per }3^{2}{\per }5$&8&CF&$61$&$.62$&$750$&$$&$134$&$.46$&$14{,}074{,}851$&$$&$0$&$.40$&$43{,}636$&$$\\%
pp16-19&\cite {bliss}&546&4641&$2^{8}{\per }3^{2}$&14&CF&$60$&$.62$&$10$&$$&$1{,}189$&$.43$&$130{,}609{,}514$&$$&$3$&$.14$&$33{,}119$&$$\\%
pp16-21&\cite {bliss}&546&4641&$2^{7}{\per }3^{3}$&12&CF&$61$&$.79$&$7$&$$&$\textbf{10{,}000}$&$$&$1{,}682{,}088{,}433$&$$&$2$&$.83$&$29{,}732$&$$\\%
pp25&\cite {moorhouse}&1302&16926&$2^{7}{\per }3{\per }5^{3}{\per }31$&2&--&$\textbf{10{,}000}$&$$&$5{,}052{,}799$&$$&$2{,}546$&$.45$&$118{,}865{,}645$&$$&$17$&$.01$&$9{,}214$&$$\\%
pp27&\cite {moorhouse}&1514&21196&$2^{3}{\per }3^{7}{\per }7$&4&--&$\textbf{10{,}000}$&$$&$4{,}435{,}071$&$$&$\textbf{10{,}000}$&$$&$818{,}906{,}445$&$$&$52$&$.82$&$30{,}004$&$$\\%
pp64&\cite {royleprivate}&8322&270465&$2^{15}{\per }3^{3}{\per }7^{2}$&8&--&$\textbf{10{,}000}$&$$&$197{,}575$&$$&$\textbf{10{,}000}$&$$&$319{,}961{,}863$&$$&$416$&$.82$&$80{,}479$&$$\\%
mz-aug2-18&\cite {bliss}&432&684&$2^{38}$&252&--&$62$&$.04$&$5{,}374{,}331$&$$&$3$&$.51$&$1{,}048{,}954$&$$&$0$&$.92$&$73{,}415$&$$\\%
mz-aug2-20&\cite {bliss}&480&760&$2^{42}$&280&--&$297$&$.56$&$23{,}593{,}421$&$$&$14$&$.80$&$4{,}194{,}764$&$$&$1$&$.10$&$77{,}379$&$$\\%
mz-aug2-22&\cite {bliss}&528&836&$2^{46}$&308&--&$1{,}407$&$.27$&$102{,}760{,}999$&$$&$64$&$.64$&$16{,}777{,}766$&$$&$1$&$.50$&$94{,}211$&$$\\%
mz-aug2-30&\cite {bliss}&720&1140&$2^{62}$&420&--&$\textbf{10{,}000}$&$$&$429{,}873{,}822$&$$&$\textbf{10{,}000}$&$$&$1{,}801{,}193{,}025$&$$&$3$&$.63$&$178{,}099$&$$\\%
mz-aug2-50&\cite {bliss}&1200&1900&$2^{102}$&700&--&$\textbf{10{,}000}$&$$&$275{,}778{,}055$&$$&$\textbf{10{,}000}$&$$&$1{,}283{,}172{,}635$&$$&$19$&$.69$&$597{,}191$&$$\\%
had-52&\cite {bliss}&208&5512&$2^{4}{\per }13$&2&CQ&$0$&$.08$&$13$&$$&$0$&$.33$&$15{,}533$&$$&$0$&$.12$&$1{,}553$&$$\\%
had-100&\cite {bliss}&400&20200&$2^{4}{\per }5^{2}$&2&CQ&$1$&$.47$&$13$&$$&$2$&$.62$&$58{,}613$&$$&$1$&$.63$&$3{,}679$&$$\\%
had-184&\cite {bliss}&736&68080&$2^{6}{\per }23$&2&CQ&$30$&$.69$&$107$&$$&$23$&$.34$&$125{,}285$&$$&$6$&$.29$&$28{,}652$&$$\\%
had-232&\cite {bliss}&928&108112&$2^{6}{\per }29$&2&CQ&$92$&$.56$&$128$&$$&$56$&$.51$&$199{,}679$&$$&$14$&$.93$&$32{,}066$&$$\\%
had-sw-32-1&\cite {bliss}&128&2112&$2^{2}$&42&CQ&$0$&$.05$&$66$&$$&$2$&$.79$&$141{,}932$&$$&$0$&$.07$&$3{,}095$&$$\\%
had-sw-88&\cite {bliss}&352&15664&$2^{2}$&132&CQ&$18$&$.75$&$140$&$$&$214$&$.73$&$3{,}645{,}512$&$$&$2$&$.34$&$12907$&$$\\%
had-sw-112&\cite {bliss}&448&25312&$2^{2}$&224&CQ&$90$&$.63$&$226$&$$&$555$&$.42$&$1{,}090{,}338$&$$&$6$&$.22$&$10{,}230$&$$\\%
had-236&\cite {bliss}&944&111864&$2$&472&CQ&$3{,}054$&$.55$&$9$&$$&$\textbf{10{,}000}$&$$&$98{,}665{,}940$&$$&$99$&$.63$&$22{,}014$&$$\\%
GenQuad-1&\cite {royleprivate}&2752&481600&$2^{8}{\per }3^{2}{\per }7^{5}$&2&--&$\textbf{10{,}000}$&$$&$5{,}989{,}083$&$$&$997$&$.82$&$1{,}291{,}833$&$$&$3$&$.14$&$30{,}339$&$$\\%
GenQuad-2&\cite {royleprivate}&7300&2693700&$2^{7}{\per }3^{10}{\per }5^{2}$&2&--&$\textbf{10{,}000}$&$$&$1{,}502{,}419$&$$&$\textbf{10{,}000}$&$$&$50{,}993{,}419$&$$&$3$&$.91$&$2{,}071$&$\hspace{.3mm}\star$\\%
Hypercube&\cite {mckayprivate}&3161&18780&$2^{5}{\per }3^{2}$&42&--&$1{,}545$&$.82$&$186{,}798$&$$&$47$&$.74$&$186{,}796$&$$&$9$&$.69$&$1{,}484$&$\hspace{.3mm}\star$\\\hline %
\end {tabular}%
\vspace*{1.6mm}
\caption{Graphs with large search tree: Automorphism Group  (\emph{nauty} invariants \cite{nauty22}: CF = cellfano2, CQ~=~cellquads; $\star:$ run with $1$-dimensional refinement)}
\label{t:res_aut}
\end{table}

\renewcommand\arraystretch{1}
\renewcommand{\tabcolsep}{1pt}
\begin{table}[t]
\begin{scriptsize}
\begin {tabular}{|c|c|r|r|c|c||c|r<{\pgfplotstableresetcolortbloverhangright }@{}l<{\pgfplotstableresetcolortbloverhangleft }|r<{\pgfplotstableresetcolortbloverhangright }@{}l<{\pgfplotstableresetcolortbloverhangleft }||r<{\pgfplotstableresetcolortbloverhangright }@{}l<{\pgfplotstableresetcolortbloverhangleft }|r<{\pgfplotstableresetcolortbloverhangright }@{}l<{\pgfplotstableresetcolortbloverhangleft }||r<{\pgfplotstableresetcolortbloverhangright }@{}l<{\pgfplotstableresetcolortbloverhangleft }|r<{\pgfplotstableresetcolortbloverhangright }|@{}r<{\pgfplotstableresetcolortbloverhangleft }|}%
\hline  & & & & \multicolumn {1}{c|}{ }& &\multicolumn {5}{c||}{{\bf\emph{nauty 2.4}}}&\multicolumn {4}{c||}{\bf \emph{bliss 0.50}}&\multicolumn {4}{c|}{\bf\emph{Traces}}\\%
Graph&Ref&V&E&$\mid \text {Aut}\mid $&Orbs&\multicolumn {1}{c}{Inv}&\multicolumn {2}{c}{Time}&\multicolumn {2}{c||}{Size}&\multicolumn {2}{c}{Time}&\multicolumn {2}{c||}{Size} &\multicolumn {2}{c}{Time}&\multicolumn {1}{c}{Grp}&\multicolumn {1}{c|}{Size}\\\hline %
ag2-16&\cite {bliss}&528&4352&$2^{14}{\per }3^2{\per }5^2{\per }17$&2&--&$0$&$.02$&$155$&$$&$0$&$.01$&$102$&$$&$0$&$.03$&$0.03$&$208$\\%
ag2-49&\cite {bliss}&4851&120050&$2^{10}{\per }3^2{\per }5^2{\per }7^6$&2&--&$0$&$.87$&$51$&$$&$0$&$.25$&$51$&$$&$0$&$.36$&$0.12$&$71$\\%
cfi-$20$&\cite {bliss}&200&300&$2^{11}$&80&--&$0$&$.02$&$146$&$$&$0$&$.02$&$259$&$$&$0$&$.01$&$0.01$&$186$\\%
cfi-$80$&\cite {bliss}&800&1200&$2^{41}$&320&--&$2$&$.79$&$1{,}125$&$$&$0$&$.07$&$1{,}999$&$$&$2$&$.73$&$0.67$&$1{,}049$\\%
difp-20-0&\cite {bliss}&8965&23082&$1$&8965&--&$1$&$.89$&$1$&$$&$0$&$.01$&$1$&$$&$0$&$.16$&$0.00$&$1$\\%
fpga-10-8&\cite {bliss}&688&1320&$2^{23}{\per }3$&519&--&$0$&$.01$&$300$&$$&$0$&$.01$&$300$&$$&$0$&$.18$&$0.14$&$242$\\%
fpga-13-11&\cite {bliss}&1500&3125&$2^{32}{\per }3^7$&1116&--&$0$&$.21$&$780$&$$&$0$&$.01$&$780$&$$&$0$&$.97$&$0.76$&$560$\\%
grid-3-20&\cite {bliss}&8000&22800&$2^{4}{\per }3$&220&--&$1$&$.33$&$5$&$$&$0$&$.03$&$5$&$$&$1$&$.40$&$0.09$&$4$\\%
grid-w-2-100&\cite {bliss}&10000&20000&$2^{7}{\per }5^4$&1&--&$12$&$.43$&$7$&$$&$0$&$.04$&$7$&$$&$1$&$.99$&$0.28$&$11$\\%
grid-w-3-20&\cite {bliss}&8000&24000&$2^{10}{\per }3{\per }5^3$&1&--&$3$&$.36$&$8$&$$&$0$&$.04$&$8$&$$&$1$&$.98$&$0.26$&$14$\\%
k-70&\cite {bliss}&70&2415&$>1.97\cdot 10^{100}$&1&--&$0$&$.01$&$2{,}485$&$$&$0$&$.01$&$2{,}485$&$$&$0$&$.88$&$0.85$&$689$\\%
k-100&\cite {bliss}&100&4950&$>9.33\cdot 10^{157}$&1&--&$0$&$.01$&$5{,}050$&$$&$0$&$.02$&$5{,}050$&$$&$7$&$.24$&$7.18$&$890$\\%
latin-30&\cite {bliss}&900&39150&$2^{6}{\per }3^3{\per }5^2$&1&--&$0$&$.10$&$21$&$$&$0$&$.07$&$51$&$$&$0$&$.07$&$0.03$&$233$\\%
lattice-30&\cite {bliss}&900&26100&$>1.40\cdot 10^{65}$&1&--&$0$&$.16$&$930$&$$&$0$&$.13$&$901$&$$&$0$&$.74$&$0.70$&$149$\\%
mz-$18$&\cite {bliss}&360&540&$2^{39}$&90&--&$0$&$.06$&$593$&$$&$0$&$.01$&$593$&$$&$0$&$.14$&$0.13$&$500$\\%
mz-$50$&\cite {bliss}&1000&1500&$2^{103}$&250&--&$\textbf{600}$&$$&$-$&$$&$0$&$.03$&$3{,}249$&$$&$4$&$.91$&$4.73$&$2{,}283$\\%
mz-aug-$22$&\cite {bliss}&440&1012&$2^{47}$&110&--&$0$&$.10$&$730$&$$&$0$&$.01$&$826$&$$&$0$&$.26$&$0.25$&$466$\\%
paley-461&\cite {bliss}&461&53015&$2{\per }5{\per }23{\per }461$&1&--&$0$&$.01$&$6$&$$&$0$&$.04$&$6$&$$&$0$&$.01$&$0.00$&$8$\\%
pg2-32&\cite {bliss}&2114&34881&$>1.09\cdot 10^{13}$&1&--&$2$&$.71$&$1{,}040$&$$&$0$&$.16$&$753$&$$&$0$&$.21$&$0.06$&$513$\\%
rnd-3-reg-3000-1&\cite {bliss}&3000&4500&$1$&3000&DI&$0$&$.69$&$1$&$$&$0$&$.23$&$3{,}001$&$$&$0$&$.67$&$0$&$1$\\%
rnd-3-reg-10000-1&\cite {bliss}&10000&15000&$1$&10000&DI&$39$&$.36$&$1$&$$&$2$&$.86$&$10{,}001$&$$&$6$&$.89$&$0$&$1$\\%
s3-3-3-3&\cite {bliss}&11076&20218&$2^{12}$&7836&--&$1$&$.07$&$91$&$$&$0$&$.03$&$91$&$$&$0$&$.92$&$0.61$&$272$\\%
urq8-5&\cite {bliss}&3906&20331&$1$&3906&--&$0$&$.12$&$1$&$$&$0$&$.01$&$1$&$$&$0$&$.06$&$0$&$1$\\\hline %
\end {tabular}%
\end{scriptsize}
\vspace*{1.6mm}
\caption{Graphs with small search tree: Canonical Form  (\emph{nauty} invariants \cite{nauty22}: DI = distances)}
\label{t:res_small}
\end{table}

\subsection{Experiments and comments}

Experiments were carried out on a Apple MacBook Pro with Intel Core i7 processor at $2.66$ GHz
and $4$ GB RAM, under gcc $4.0$.

Graphs are selected from the library of benchmarks which is attached to the \emph{bliss} distribution (\cite{bliss}), 
with the addition of some very hard graphs; these are available in DIMACS format at the \emph{Traces} web page. 
Concerning the benchmark 
families of graphs which are not presented here, either they display trivial results (small differences among all of the tools), 
or they reveal results similar to some of the presented classes.

With reference to the classification in the \emph{bliss} library we have selected graphs from the following families:
\begin{compactitem}
\item[-] affine and projective geometries: graphs \textsf{ag2-x}, \textsf{pg2-32};
\item[-] Cai-F\"urer-Immerman construction: graphs \textsf{cfi-x};
\item[-] constraint satisfaction problems: graphs \textsf{difp-20-0}, \textsf{fpga-x-y}, \textsf{s3-3-3-3}, \textsf{urq8-5};
\item[-] Hadamard matrices: graphs \textsf{had-x}, \textsf{had-sw-x} (with some switching operations);
\item[-] Miyazaki constructions: graphs \textsf{mz-x}, \textsf{mz-aug-x}, \textsf{mz-aug2-x};
\item[-] projective planes: graphs \textsf{ppx}: some of them are from \cite{moorhouse} or from Gordon Royle~\cite{royleprivate};
\item[-] other graphs of combinatorial origin: graphs \textsf{GenQuad-x} and \textsf{Hypercube} are from Gordon Royle \cite{royleprivate} and 
Brendan McKay \cite{mckayprivate};
\item[-] random regular graphs: graphs \textsf{rnd-3-reg-x-y};
\item[-] strongly regular graphs: graphs \textsf{latin-x}, \textsf{latin-sw-x-y}, \textsf{lattice-30}, \textsf{sts-x}, \textsf{sts-sw-x-y};
\item[-] complete graphs: graphs \textsf{k-x};
\item[-] grid graphs: graphs \textsf{grid-x-y}, \textsf{grid-w-x-y}.
\end{compactitem}

For each experiment the following information is reported in Tables \ref{t:res_can} 
(canonical labeling of graphs with large search tree),  
\ref{t:res_aut} (automorphism group computation for graphs with large search tree) and \ref{t:res_small}
(canonical labeling of graphs with small search tree):
the name of the graph and a reference to it, the number of its vertices and edges; the size of the automorphism group
of the graph and the number of its orbits; for \emph{nauty}, the vertex-invariant used; for all the tools considered, 
the execution time (in seconds) and the size of the associated search space. When an experiment is interrupted
(after a reported number of seconds shown in bold face), the size of the already computed portion of the search space is displayed.

$2$-dimensional refinement is used for the graphs reported in Tables  \ref{t:res_can},\ref{t:res_aut}, while
$1$-dimen-sional refinement is used for the graphs in Table \ref{t:res_small}. Diagrams in Figures \ref{f:pag1} and \ref{f:pag2} 
present comparison among the considered tools for some classes of graphs.  

\subsubsection{Tables \ref{t:res_can},\ref{t:res_aut}: graphs with large search space}
For all the graphs considered, \emph{Traces} exhibits a drastic decrease of the size of their search space, with clear
consequences for computation time.The gain in performance of \emph{Traces} with respect to \emph{nauty} and \emph{bliss} is considerable
for all graphs of combinatorial origin. Several classes of graphs which cannot be efficiently treated by \emph{nauty} and \emph{bliss}
are handled by \emph{Traces} in a few seconds.

It turns out that the hardest instances are graphs with small automorphism group,
such as some Hadamard graphs. In particular, critical examples for \emph{Traces} with
$1$-dimensional refinement are from the \textsf{had-sw} family. The contrast
with the efficiency of the $2$-dimensional refinement suggests that the comparison
between refinement traces has to be improved in the case of $1$-dimensional refinement.

\begin{figure}[htb]
\begin{center}
\begin{tikzpicture}[transform shape, scale=.70]
\begin{semilogyaxis}[
ymin=0,  ymax=5000, 
enlargelimits=0.1, 
xlabel=Automorphism group size, 
ylabel=Time, 
grid=major, 
legend style={at={(0.5,1.118)},
anchor=north,legend columns=-1}]
\Bliss
coordinates {
( 3840, 618.94 )
( 12288, 109.77 )
( 18432, 3196.47 )
( 147456, 576.33 )
( 884736, 1.87 )
( 921600, 200.03 )
( 258048, 300.81 )
( 55296, 1705.60 )
( 92160, 65.79 )
( 2304, 835.73 )
( 3456, 4471.48 )
};
\addlegendentry{\emph{bliss}}
\Nauty
coordinates {
( 3840, 59.66 )
( 12288, 68.39 )
( 18432, 60.72 )
( 147456, 61.63 )
( 884736, 61.01 )
( 921600, 62.55 )
( 258048, 61.29 )
( 55296, 61.41 )
( 92160, 60.52 )
( 2304, 59.62 )
( 3456, 59.63 )
};
\addlegendentry{\emph{nauty}} 
\TracesTwoD
coordinates {
( 3840, 44.77 )
( 12288, 11.49 )
( 18432, 12.93 )
( 147456, 1.18 )
( 884736, 0.32 )
( 921600, 0.31 )
( 258048, 0.76 )
( 55296, 2.80 )
( 92160, 1.77 )
( 2304, 55.95 )
( 3456, 38.25 )
};
\addlegendentry{\emph{Traces$\,_{2-\mathrm{dim}}$}}
\TracesOneD
coordinates {
( 3840, 199.27 )
( 12288, 73.78 )
( 18432, 38.28 )
( 147456,2.77 )
( 884736, 0.42 )
( 921600, 0.42 )
( 258048, 1.34 )
( 55296, 6.58 )
( 92160, 3.64 )
( 2304, 438.06 )
( 3456, 273.43 )
};
\addlegendentry{\emph{Traces$\,_{1-\mathrm{dim}}$}}
\end{semilogyaxis} 
\end{tikzpicture}%
\hspace{1cm}%
\begin{tikzpicture}[transform shape, scale=.70]
\begin{semilogyaxis}[
enlargelimits=0.1, 
xlabel=Permutation, 
ylabel=Search space size (refinements), 
grid=major, 
legend style={at={(0.5,1.118)},
anchor=north,legend columns=-1}]
\Bliss
coordinates{(0,199505) (1,2322001) (2,1439210) (3,17156953) (4,3512661) (5,6289735)};
\addlegendentry{\emph{bliss}}
\TracesTwoD
coordinates{(0,4561) (1,11010) (2,4922) (3,5090) (4,14376) (5,8115)};
\addlegendentry{\emph{Traces$\,_{2-\mathrm{dim}}$}}
\TracesOneD
coordinates{(0,10954) (1,21333) (2,31381) (3,10927) (4,10689) (5,16835)};
\addlegendentry{\emph{Traces$\,_{1-\mathrm{dim}}$}}
\end{semilogyaxis} 
\end{tikzpicture}%
\caption{Non-Desarguesian projective planes of order $16$ (left); relabelings of \textsf{PP16-8} (right)}
\label{f:ppandstab}
\end{center}
\end{figure}

The reader can verify (see also Figure \ref{f:pag2} (bottom)) an exponential contraction of the search space of 
\emph{Traces} with respect to those of \emph{nauty} and \emph{bliss} 
in the case of Miyazaki's sequence \textsf{mz-aug2-x}. Now, Miyazaki's graphs are 
carefully tuned (their labeling, too) to cause \emph{nauty} (and therefore \emph{bliss}) as much trouble as possible, 
but other more ``natural'' classes of graphs which are intractable for \emph{nauty} and \emph{bliss}, such as unions of
non isomorphic strongly regular graphs with the same parameters, are efficiently treated by \emph{Traces}. 
At present, we do not know of any class of graphs forcing \emph{Traces} to exhibit a proven exponential behavior.

Finally, we observe (see Table \ref{t:res_aut}) that \emph{Traces}' performance is almost always better
in the automorphism group computation mode compared to the canonical labeling mode. 

\subsubsection{Table \ref{t:res_small}: graphs with small search space.}
Graphs exhibiting a small search space (with respect to the size of their vertex set) turn out to have either a large automorphism
group or a trivial one. In the first case, the search space is massively pruned by automorphisms, in the second case just
a few individualization steps are needed to obtain discrete partitions. These are the most favorable situations
for the individualization-refinement technique.

Still, \emph{Traces} is able to reduce both the depth and size of the search space, and it almost always displays
better performances than \emph{nauty}. If we disregard the cases where \emph{nauty} uses vertex invariants, 
the sizes of search trees of \emph{nauty} and \emph{bliss} are always similar, often exactly the same. Therefore,
it is reasonable to assume that the difference between their performances is due to the efficiency of \emph{bliss} in 
handling sparse graphs, and to expect an improvement in \emph{Traces}' performances when
suitable data structures are adopted. 

The time spent by \emph{Traces} via the Schreier-Sims algorithm in group computation 
(reported in Table  \ref{t:res_small}) becomes significant when dealing with very large 
automorphism groups: this is evident in the case of complete graphs. 

\tikzstyle{every pin}=[fill=white,
draw=black,
font=\footnotesize]
\begin{figure}[ht]
\begin{center}
\input{cfi}

\vspace{.5cm}
\input{grid}

\vspace{.5cm}
\input{mz}
\caption{}
\label{f:pag2}
\end{center}
\end{figure}%

\subsubsection{Remarks about some particular experiments}

\emph{Traces} seems to have a stable behavior on different instances of graphs in the same family
and also on different representations of the same graph, as shown in Figure \ref{f:ppandstab}.
In particular, Figure \ref{f:ppandstab} (left) compares the execution time of \emph{Traces} and \emph{bliss} with respect to the size
of the automorphism groups of graphs in the family \textsf{pp-16} (projective planes of order $16$), clearly showing 
that graphs with larger group are more efficiently treated by \emph{Traces}. This is what we expect from tools
based on pruning by automorphism.

Figure \ref{f:ppandstab} (right) compares the sizes of search spaces coming from random permutations 
of vertices of  \textsf{pp-16-8}. It comes out that \emph{Traces} has a stronger ability of capturing the 
structure of the graph, thus abstracting from its representation.
A further evidence of such claim can be also deduced from the series \textsf{cfi-x}, where, though not reported 
into the tables, the depth of \emph{Traces}' search space is equal 
to the main parameter introduced by Cai, F\"urer and Immerman in their construction \cite{CaiFI92}.

Incidence graphs of projective planes are considered among the hardest examples for practical graph
isomorphism testing, since they exhibit a high degree of regularity, whilst they can have a rather small
automorphism group. 
The best known algorithm for such graphs is due to Miller \cite{miller78} and is based on a theorem
by Bruck \cite{bruck} about the order of subplanes into projective planes. It is interesting to mention
that \emph{Traces} exactly mimics the behavior of Miller's algorithm, thanks to the correct response
of the target cell selector, without the need of any ad hoc modification.
In fact, at every iteration an individualization selects a vertex corresponding to a point of the plane which is
not collinear with the previously selected ones. The refinement (either $1$-, or $2$-dimensional)
yields a partition whose singleton vertices constitute a subplane of the input plane. This seems to be a further
evidence of the fact that \emph{Traces} captures relevant structural properties of such graphs.
In addition, Miller's construction does not consider, as \emph{Traces} does, the presence of automorphisms; 
therefore \emph{Traces} always runs below the theoretical bound established in \cite{miller78}. 
We are currently investigating whether automorphism detection may turn Miller's algorithm, 
which has a subexponential time complexity, into a polynomial one.

Finally, we have experimented with the possibility of running \emph{Traces} when the automorphism group of the input
graph is known in advance. The group can be computed more efficiently than canonical labeling in \emph{Traces}, 
as shown comparing Tables \ref{t:res_can} and \ref{t:res_aut}. Interesting results can be obtained for
large and highly symmetric graphs. For instance, the graph \textsf{GenQuad-2}, which is the collinearity graph 
of a generalized quadrangle of order $(9,81)$, can be canonized
in $7$ seconds (instead of more than $5000$!) when its automorphism group, which can be computed in $4$
seconds with simple refinement, is known. This huge difference is due to the use of information coming from the 
group structure during the refinement process.

\begin{figure}[p]
\begin{center}
\input{had}

\vspace{.5cm}
\input{hadrefs}

\vspace{.5cm}
\input{lat-sw}

\caption{}
\label{f:pag1}
\end{center}
\end{figure}

\bibliographystyle{plain}
\bibliography{Paper}		

\end{document}